\newif\ifuseprd
\newif\iftoomuchdetail
\useprdfalse 
\toomuchdetailfalse

\newif\ifatitp

\ifuseprd 
\documentclass[preprint,tightenlines,floats,prd,eqsecnum,nobibnotes%
\ifatitp,superscriptaddress\fi,nofootinbib]{revtex4}
\else
\documentclass[final,12pt,letterpaper]{JHEP}
\fi 
\usepackage{amsmath,amssymb}

\allowdisplaybreaks[3]
\newcommand\skipthis[1]{{}}

\makeatletter
\def\@strike{\relax\leavevmode
  \ifmmode
    \expandafter\mathpalette\expandafter\math@strike
  \else
    \expandafter\make@strike
  \fi}
\def\math@strike#1#2{%
  \setbox\z@\hbox{$\m@th#1{#2}$}\fin@strike}
\def\make@strike#1{%
  \setbox\z@\hbox{\color@begingroup#1\color@endgroup}\fin@strike}
\def\fin@strike{%
  \@tempdima\dp\z@
  \@tempdimb\ht\z@
  \lower\@tempdima\hbox{\strike@start}%
  \box\z@
  \raise\@tempdimb\hbox{\strike@end}}
\def\strike@start{\special{ps: %
    currentpoint /starty exch def /startx exch def}}
\def\strike@end{
}
\newcommand\fs{\protect\@strike}

\let\oldAE\AE
\renewcommand\AE{{\ifmmode{\text{\it\oldAE}}\else{\oldAE}\fi}}

\newcommand\wick[1]{{\ensuremath{\widehat{{#1}}}}}

\newcommand\mathone{{\rlap{\kern .25em l}1}}
\newcommand\one{{\ifmmode{\text{\mathone}}\else{\mathone}\fi}}

\newcommand\ct[1]{{\ifuseprd{\em{#1}},\else{\sf {#1}},\fi}}
\newcommand\bt[1]{{\em {#1}},}

\newcommand\phepth[1]{{\tt [\hepth{#1}]}}

\DeclareMathOperator{\real}{Re}

\DeclareMathOperator{\diag}{diag}

\chardef\til=`~

\newcommand\half{{\ensuremath{\frac{1}{2}}}}
\newcommand\thalf{{\ensuremath{\tfrac{1}{2}}}}
\newcommand\p{\ensuremath{\partial}}

\newcommand\abs[1]{\ensuremath{\left\lvert{#1}\right\rvert}}

\newcommand\transpose{{\ensuremath{\text{\sf T}}}}

\newcommand\order[1]{{\ensuremath{{\mathcal O}({#1})}}}
\newcommand\vev[1]{{\ensuremath{\left\langle{#1}\right\rangle}}}
\newcommand\anti[2]{\ensuremath{\left\{{#1},{#2}\right\}}}
\newcommand\com[2]{\ensuremath{\left[{#1},{#2}\right]}}
\DeclareMathOperator{\Tr}{Tr}

\newcommand\sfrac[2]{\ensuremath{{{#1}}/{({#2})}}} 

\newcommand\apr{{\ensuremath{{\alpha'}}}}

\newcommand\lam{\lambda}

\newcommand\sT{{\ensuremath{{\mathcal T}}}}
\newcommand\ha{{\half}}

\newcommand\ch{{\mathcal C\/}}

\newcommand\cM{{\ensuremath{{\mathcal{M}}}}}
\newcommand\cC{{\ensuremath{{\mathcal{C}}}}}

\newcommand\co[1]{{\ensuremath{{\iota_{{#1}}}}}}
\newcommand\Pb{{\ensuremath{{\mathcal{P}}}}}

\DeclareMathOperator{\Pf}{Pf}
\DeclareMathOperator{\STr}{STr}
\newcommand\swz{{\ensuremath{S_{\text{WZ}}}}}

\newif\ifwz
\wzfalse
\newif\ifrr
\rrfalse
\newcommand\WZ{{\ifwz{WZ}\else{Wess-Zumino (WZ)}\fi\global\wztrue}}
\newcommand\RR{{\ifrr{RR}\else{Ramond-Ramond (RR)}\fi\global\rrtrue}}

\providecommand\putabstract[1]{\ifuseprd\begin{abstract} {#1} \end{abstract}%
                           \else \abstract{{#1}} \fi}
\providecommand\plb[3]{{Phys.\ Lett.\ B {\bf {#1}}, {#3} ({#2})}}
\providecommand\npb[3]{{Nucl.\ Phys.\ {\bf B{#1}}, {#3} ({#2})}}
\providecommand\jhep[3]{{J.\ High Energy Phys.\ {\bf #1}, {#3} ({#2})}}

\providecommand\ptp[3]{{\ifuseprd\else\begingroup\em\fi Prog.\ Theor\ Phys.\ %
     \ifuseprd\else\endgroup\fi {\bf {#1}}\ifuseprd, {#3} ({#2})\else%
     \ ({#2}) {#3}\fi}}
\providecommand\cqg[3]{{\ifuseprd\else\begingroup\em\fi Class.\ and Quant.\ %
     Grav.\ %
     \ifuseprd\else\endgroup\fi {\bf {#1}}\ifuseprd, {#3} ({#2})\else%
     \ ({#2}) {#3}\fi}}
\providecommand\cjm[3]{{\ifuseprd\else\begingroup\em\fi Canad.\ J.\ Math.\ %
     \ifuseprd\else\endgroup\fi {\bf {#1}}\ifuseprd, {#3} ({#2})\else%
     \ ({#2}) {#3}\fi}}
\newcommand\citeprd[3]{{\ifuseprd{Phys.\ Rev.\ D {\bf {#1}}, {#3} ({#2})}%
                        \else{\prd{#1}{#2}{#3}}\fi}}
\providecommand\hepth[1]{{\tt hep-th/{#1}}}
\newenvironment{smaleq}{\ifuseprd\else\small\fi}{}

\ifuseprd
\begin{document} 
\fi 

\title{
{\ifuseprd\else\Huge\fi$\ast$-Trek III:} \\ The Search for
Ramond-Ramond Couplings}
\ifuseprd
\author{Hong Liu}\email{liu@physics.rutgers.edu}
\affiliation{New High Energy Theory Center,
       Rutgers University;
       126 Frelinghuysen Road;
       Piscataway, NJ \ 08854}
\author{Jeremy Michelson}\email{jeremy@physics.rutgers.edu}
\affiliation{New High Energy Theory Center,
       Rutgers University;
       126 Frelinghuysen Road;
       Piscataway, NJ \ 08854}
\ifatitp
\affiliation{Institute for Theoretical Physics,
University of California;
Santa Barbara, CA \ 93106-4030}
\fi
\else 
\author{Hong Liu\ifatitp$^1$\fi\thanks{\tt liu@physics.rutgers.edu} \ 
and
Jeremy Michelson\ifatitp$^{1,2}$\fi\thanks{\tt jeremy@physics.rutgers.edu} \\
{\ifatitp$^1$\fi}New High Energy Theory Center \\
Rutgers University \\
126 Frelinghuysen Road \\
Piscataway, NJ \ 08854 \ USA 
\ifatitp\\
\vspace{\baselineskip}
$^2$Institute for Theoretical Physics \\
University of California \\
Santa Barbara, CA \ 93106-4030
\fi} 
\fi 

\putabstract{
We give a%
\iftoomuchdetail%
n overly%
\fi\ %
detailed discussion of the disk amplitudes with one closed
string insertion, which we used to construct the supergravity
couplings of noncommutative D-branes to the RR potentials, given
in~\hepth{0104139}.  We prove the inclusion of Elliott's formula, the
integer-valued modification of the noncommutative Chern character, to
all orders in the gauge field.  We also give a detailed comparison
between the form of the result in which Elliott's formula is manifest,
and the form expressed in Matrix model variables.
}

\preprint{RUNHETC-2001-20\ifuseprd,~\else\\\fi 
   {\tt hep-th/0107172}}

\ifuseprd
\maketitle
\else
\begin{document}
\fi 

\wzfalse\rrfalse

\section{Introduction} \label{sec:intro}

In the past few years, great insights have been gained  by studying the 
couplings of D-branes to bulk closed string modes.
Examples include  gauge theory dynamics, black holes and the AdS/CFT 
correspondence. In particular studying \RR\ couplings~\cite{miao,ghm}
has yielded the ``branes-within-branes'' phenomenon~\cite{bwbw,bwbd}, K-theory 
descriptions of D-branes~\cite{mm,witk}, Myers 
effects\cite{myers,taylor}
and better understandings of anomalies.

When we turn on a constant  background  Neveu-Schwarz 
$B$-field in the presence of a D-brane, the low energy worldvolume
theory of the D-brane becomes noncommutative~\cite{cds,dh,sw} (for a
review see~\cite{dnrev}). 
In this paper we continue
to study couplings of noncommutative D-branes
to spacetime gravity fields.
In previous papers~\cite{liu,lm3}, we examined the  couplings
to the fluctuations of the closed string metric $g_{\mu\nu}$, dilaton 
and the $B$-field.  In this paper we will examine the couplings 
to \RR\ potentials.  We have already presented our results
in~\cite{lm4}.  The purpose of this paper is to provide a detailed
derivation of the results.
Closely related discussions of the couplings of 
noncommutative D-branes to background closed string modes have appeared
in~\cite{dastrivedi,ooguri,oo,mukhi,mukhi2,garousi1,dasmu,dhar}. 

In the presence of a constant
Neveu-Schwarz $B$-field, the worldsheet open string 
propagator is
\begin{equation}
\vev{X^{\mu} (\tau) X^{\nu} (\tau')} = - \apr G^{\mu \nu} \log (\tau 
-\tau')^2 + \frac{i}{2} \theta^{\mu \nu} \epsilon (\tau - \tau')
\end{equation}
where the open string metric $G$ and the noncommutative parameter $\theta$ 
are given in terms of the constant background closed string parameters by
\begin{subequations} \label{defGt}
\begin{align} \label{defG}
G_{\mu \nu} & = g_{\mu \nu} - (B g^{-1} B)_{\mu \nu}, \\
\label{deftheta}
\theta^{\mu\nu} &= - (2 \pi \alpha')  
\left ( \frac{1}{g+B} B  \frac{1}{g-B} \right)^{\mu \nu} \ .
\end{align}
\end{subequations}
In particular the mass shell condition for open string modes 
is $k_{\mu} G^{\mu \nu} k_{\nu} \in \frac{{\mathbb{Z}}_+}{2 \apr}$.
Thus the limit $\apr \rightarrow 0$, with
$\theta$ and $G$ fixed, yields a noncommutative gauge theory with 
noncommutative parameter $\theta$. 
Here we shall be interested  in the couplings between the RR modes 
and the noncommutative gauge modes to leading order in $\apr$. 
They are to be extracted from the on-shell disk amplitudes.

Since the open string metric $G$ and the closed string metric $g$ 
have different scaling limits with respect to $\apr$ with 
fixed $\theta$, the low energy limit on the brane in 
terms of the open string metric no longer corresponds to the low
energy limit
in terms of the closed string metric in the bulk. Thus the low energy 
modes on the brane generically excite finite momentum \RR\ fields in
the bulk,
even to lowest order in $\apr$. In the D-brane worldvolume theory
this is reflected in the presence of an open Wilson 
line~\cite{iikk,ambjorn,dasrey,ghi} with an appropriate integration 
prescription~\cite{liu}. 
Open Wilson lines are also necessary in order to have off-shell gauge 
invariance in the noncommutative gauge theory%
\footnote{For a discussion of noncommutative gauge invariant operators from 
a different perspective, see~\cite{bl}. See also~\cite{kiemrey} for a 
discussion of open Wilson lines in noncommutative scalar field theories.}.  
An open Wilson line may be
considered as the object which transforms an object in 
the open string algebra to an element of the closed string algebra. 
In explicit amplitude calculations with a finite number of 
external open string modes, the Wilson lines
manifest themselves in the form
of $n$-ary  $\ast_n$ operations~\cite{garousi,lm2,mehen,liu}.

Microscopically, the appearance of the Wilson line may be understood 
from the fact that in noncommutative field theories, the elementary quanta are 
dipoles whose lengths are proportional to their transverse 
momenta~\cite{bs,sj,yin}. These dipoles interact by
splitting and joining their ends. When a closed string mode scatters 
off open string modes on a D-brane, 
all external open string modes join together to form a macroscopic 
open Wilson line, which then couples to the closed 
string mode. 
The Wilson line can thus be viewed as the spacetime image of the 
worldsheet boundary.

The leading \RR\ couplings  we find are topological in nature and contain
Elliott's 
formula involving the noncommutative Chern character~\cite{elliott,rieffel}. 
In particular for D-branes which are described by topologically
nontrivial configurations of the world-volume gauge theory of higher 
dimensional branes, their \RR\ charges are given precisely by Elliott's 
formula. This  result fits in well with the
expectations of~\cite{schwarz,schwarzre}, based on the K-theory 
of noncommutative tori.
The results we find  can also be written in various other forms. Each form is 
convenient for understanding certain physical aspects, such as
the relation with noncommutative K-theory; the Matrix model; and
the Seiberg-Witten map between different descriptions.

As pointed out in~\cite{sw}, there are different descriptions 
of the D-brane dynamics parameterized by a noncommutative parameter 
$\theta$ or an open string two-form background $\Phi$, via
\begin{equation} \label{swp}
\frac{1}{g + B} = \frac{1}{G + \Phi} + \frac{\theta}{2 \pi \apr}.
\end{equation}
The results we find from the on-shell amplitudes correspond to the 
description $\Phi =0$ with $\theta$ given by~\eqref{deftheta}.
Different descriptions have different open string metrics $G$ and
different $\apr$ expansions.
Since the mass shell conditions for massive 
open string states are defined in terms of the open string metric $G$
in the $\Phi =0$ description~\eqref{defG}, the leading order results
(in terms of $\apr$) for other values of $\Phi$ generally involve the 
contributions of massive open string modes.
The \RR\ couplings in other descriptions, and the relations 
between them, are discussed in~\cite{lm4}. 

We believe the results presented here and in~\cite{liu,lm3,lm4} 
give the couplings from which one can read off the CFT operators
corresponding to
various supergravity fields 
in the noncommutative 
version of the $AdS$/CFT correspondence~\cite{hi,mr}.
These results should also provide clues for understanding the closed string
modes in the open string field theory.   

This paper is organized as follows.
In section~\ref{sec:prelim} we set our conventions and derive the
basic correlation functions that we use to compute the amplitude.  One
interesting
result here is the boundary
condition for the spacetime spinor fields on the worldsheet.  An
explicit derivation of this boundary condition is given in
appendix~\ref{app:m}.  In section~\ref{sec:amp}, we compute the
one and two external open string amplitudes.  The corresponding action
is given in~\ref{sec:wz} and its extension to all orders in the gauge
field is proposed.  In section~\ref{sec:mm}  we present the result 
in the Matrix model language, which has a rather simple and instructive 
form. It agrees with  results from the Matrix model~\cite{oo,mukhi2} 
in the infinite $B$ limit.
Finally, in section~\ref{sec:wepb}, we extract, from arbitrary order diagrams,
all but the ``nonabelian'' terms in the action---that is, interactions
in field strengths and covariant derivatives---thereby providing
another strong check of our results.  We conclude in
section~\ref{sec:conc}.

We have several appendices containing additional details.  In
appendix~\ref{app:C*C} we discuss the \RR\ vertex operator in the 
picture $(-1/2,-3/2)$ in detail.
The worldsheet boundary conditions for the spacetime spinor fields is
derived in appendix~\ref{app:m}.  Our results depend on a somewhat complicated
$\Gamma$-matrix trace; this is derived in
detail
in appendix~\ref{app:trace}.  The amplitudes with two external open
strings require some integrations over vertex operator positions; the
exact formulas are listed in appendix~\ref{sec:ints}.
Finally, we explicitly show how our amplitude leads to the proposed
action, in appendix~\ref{sec:verify}.  In appendix~\ref{sec:exps}, we
explicitly expand out the action through quadratic order, listing all
twelve terms.

\section{Setting Up the Computation} \label{sec:prelim}

We will use $M,N = 0,1,\dots,9$ to denote the spacetime indices;
$\mu, \nu = 0,1,\dots,p$ to denote the worldvolume directions of 
a D-brane; and $i,j=p+1,\dots,9$ for the directions transverse to the brane.
In addition to
the relations~\eqref{defGt},
there is a relation between the closed and open string couplings,
$g_s$ and $G_s$.  We have
\begin{align} \label{ocmetric}
G^{-1} + \frac{\theta}{2 \pi \apr} = \frac{1}{g + B},  \\
\label{occoup}
G_s = g_s \left( \frac{\det G}{\det g} \right)^{\frac{1}{4}}.
\end{align}
We assume that $B$ lies only in Neumann
directions 
(that is, along the $D$-brane)
and
that $g_{M N}$
vanishes for mixed Neumann/Dirichlet directions.
Therefore equation~\eqref{ocmetric}
applies to all spacetime directions; in particular, $G_{ij} = g_{ij}$ and 
$\theta^{ij} = 0$.

We will put the \RR\ vertex operator in the ($-1/2,-3/2$) picture;
this will soak up all the superghost zero modes, and then we can put
all the open string vertex operators in the $0$-picture, thereby
allowing us to treat them symmetrically.  Also, the ($-1/2,-3/2$)
picture involves the \RR\ potentials rather than the field strengths,
thereby making it more natural for the purpose of finding the
couplings of D-branes. We shall take the worldsheet to be the upper 
half-plane and  use the doubling trick to extend the upper half-plane to the 
full complex plane.

We are thus interested in computing the disk amplitude 
\begin{equation} \label{namp}
{\mathcal A}_n  =  
\left[\prod_{a=2}^{n} \int_{-\infty}^{\infty} dy_a \right] 
\vev{V^{-1/2,-3/2}_{\text{RR}} (q; i) \; 
 V_O^0 (a_1, k_1; 0) \; \prod_{a=2}^{n} V_O^0 (a_a, k_a; y_a)} 
\end{equation}
where $V^{-1/2,-3/2}_{\text{RR}}$ and $V_O^0$ are vertex operators for
the massless
\RR\ potentials and open string modes (gauge bosons and/or
transverse scalar fields) respectively.
We have used  the $SL(2,{\mathbb{R}})$-invariance of 
the upper half-plane to fix the \RR\ vertex operators at $z=i$ and one of 
the open string vertex operators at the origin.
The gauge bosons and the transverse scalar fields have momenta $k_{a\mu}$ 
and polarizations $a_{aM}$, with $a,b,\dots$
labeling the open string mode; the 
massless closed string mode has momentum $q$.
Obviously, $k_a$ are purely longitudinal while $q$ can also have
transverse components because the D-brane breaks translational invariance.
Specifically,
momentum conservation requires $q_\parallel = -k\equiv -\sum_a k_a$, where 
$q_{\parallel}$ and $q_{\perp}$  denote the components of $q$ 
parallel and perpendicular to the brane.

\subsection{Vertex Operators}

As discussed in detail
in~\cite{divech}, there are many ways of writing 
the ($-1/2,-3/2$) \RR\ vertex operator.  In appendix~\ref{app:C*C},
we give additional details regarding the choice given here.
We use
\begin{align}
\label{vclose}
V^{-1/2, -3/2}_{\text{RR}} & =  \lambda g_c e^{- \phi/2 - 3\tilde{\phi}/2}
\Theta \ch \frac{\one-\Gamma^{11}}{2} \fs{C}  \tilde{\Theta} e^{i q
\cdot X}, \\
\label{vopen}
V^0_O  & =
a_{M} \, (i \dot{X}^{M}
+  2 \apr k \cdot \Psi \Psi^{M} ) e^{i k \cdot X},
\end{align}
where $C = \sum_n C^{(n)}$ is a sum of {\em all\/} the \RR\ gauge potentials; 
and $n$ is odd (even) for Type
IIA(B). 
We use a Feynman-slash notation,
\begin{equation} \label{fs}
\fs{C^{(n)}} = \frac{1}{n!} C^{(n)}_{M_1\dots M_n}
\Gamma^{M_1\dots M_n},
\end{equation}
where, of course, $\Gamma^{M_1\dots M_n}$ are antisymmetrized
$\Gamma$-matrices of Spin(1,9), with unit weight and the convention
$\anti{\Gamma^M}{\Gamma^N} = 2g^{MN}$.  Note the appearance of the
closed string metric here.  The slash is
distributive over addition.  $\Theta$ and
$\Tilde{\Theta}$ are the left- and right-moving  spin operators, 
which can be related to the worldsheet fermions $\psi$ and $\tilde{\psi}$
via bosonization.%
\footnote{Schematically, if $\psi \sim e^{iH}$ then $\Theta \sim e^{i
s\cdot H}$ where $s=\pm \half$.\label{ft:bos}} \ 
We will usually suppress spinor indices, $A,B,\cdots$.
The appearance of $\Gamma^{11} = \Gamma^0\cdots \Gamma^9$
in~\eqref{vclose} enforces the chirality of the spacetime
spinors as required by the GSO projections; this may seem
to differ from
the usual chirality conditions because the picture-changing operator
changes the chirality of $\tilde{\Theta}$
as compared to the $(-1/2,-1/2)$-picture vertex
operator involving the field strengths. 
The charge conjugation matrix is denoted $\ch$, and $\lambda$ is 
a normalization constant to be fixed later.

In equation~\eqref{vopen}, the overdot is understood to be
the tangential derivative along the worldsheet boundary 
for Neumann directions (gauge bosons), and the normal derivative  for 
Dirichlet directions (transverse scalar fields).  We have absorbed a 
factor of $g_{YM}$ into the polarization $a_M$.  
Also, we will suppress Chan-Paton factors from the formulas
for notational simplicity, although we will comment on them when appropriate.

In the following, we will largely suppress the superghosts $\phi$, 
$\tilde{\phi}$ and the $bc$ ghosts, noting only that they have the
same standard OPEs and correlators as when $B=0$.

Note that in equations~\eqref{vclose} and~\eqref{vopen}
\begin{equation} \label{defym}
g_{YM}^2  = (2 \pi)^{(p-2)} G_s \apr^{\frac{p-3}{2}}, \qquad
g_c = \frac{\kappa_{10}}{2 \pi} = \sqrt{\pi} g_s^2 (2 \pi \apr)^2,
\end{equation}
where the open and closed string couplings $G_s$ and $g_s$ are related 
by~\eqref{occoup}. 
The overall normalization constant for the disk amplitudes is given 
by~\cite{jp}
\begin{equation} \label{norm}
C_{D_2} = \frac{1}{2 g_{YM}^2 \apr^2} \sqrt{-\det G},
\end{equation}
and the brane tension is related to the Yang-Mills coupling constant $g_{YM}$ 
by
\begin{equation} \label{tension}
T_p = \frac{1}{(2 \pi)^{p} g_s \apr^{\frac{p+1}{2}}},\qquad
T_p \sqrt{-\det (g + B)} = 
\frac{1}{g_{YM}^2 (2 \pi \apr)^2 }\sqrt{-\det G}.
\end{equation}

The boundary conditions at $\bar{z} =z$ for the worldsheet fields are 
\begin{align} 
\bar{\partial} X^{M} (\bar{z}) & = 
\left(\frac{1}{g-B} (g+B) D \right)^{M}{_N} \, 
\partial X^{N} (z), \\
\label{fermb}
\tilde{\psi}^{M} (\bar{z}) & =  
\left(\frac{1}{g-B} (g+B) D \right)^{M}{_N} \; \psi^{N} (z).
\end{align}
where  $D = \diag (1,\dots,1,-1,\dots,-1)$, i.e. the identity
matrix in the Neumann directions and minus the identity
in the Dirichlet directions. 
The boundary fermion $\Psi^M$,
\begin{equation} \label{defPsi}
\Psi^{M}  = \left(\frac{1}{g-B} g \right)^{M}{_N} \; \psi^{N},
\end{equation}
is the open string supersymmetric partner of $X^M$ that
lives on the worldsheet boundary.
It thus appears in the open string vertex operator~\eqref{vopen}.

Similarly, the spin operators  $\Theta$ and $\tilde{\Theta}$ are related 
via the boundary conditions.  That is, there exists a matrix $M$ so that
at $z=\bar{z}$,
\begin{equation} \label{taft2}
\tilde{\Theta}(\bar{z}) = M \Theta(z).
\end{equation}
From~\eqref{fermb} and the consistency of the OPEs
one can show that for a D$p$-brane with $B$-field background,
\begin{equation} \label{mforrl}
M =  \frac{\sqrt{-\det g}}{\sqrt{-\det{(g+B)}}} 
\AE\left(\fs{B}\right) \Gamma^0 \cdots
\Gamma^{p} \begin{cases} \one, \quad& \text{type IIA} \\
                                   \Gamma^{11}, & \text{type IIB}. \end{cases}
\end{equation}
up to a sign convention. A similar formula appeared 
in~\cite{callan} from analyzing the boundary states in Type I theory, 
and we have followed
that paper in employing the ``antisymmetrized exponential,'' $\AE(\fs{B})$.
The equation~\eqref{mforrl} is also familiar 
as its eigenspinors give the unbroken supersymmetries
for D-branes in a background field in 
supergravity (see e.g.~\cite{berg}).  
Note that $\AE(\fs{B})$, as defined in~\cite{callan}, is the
exponential of $\fs{B}$, but with the $\Gamma$ matrices totally
antisymmetrized.  (This is equivalent to $\fs{e^B}$, where wedge
products are understood in the definition of the exponential, and
where, as in~\eqref{fs}, the Feynman slash of a sum is the sum of
Feynman slashes.) In appendix~\ref{app:m} we give a worldsheet 
derivation of~\eqref{mforrl} 
along with a more explicit expression for $\AE(\fs{B})$.

\subsection{Basic Worldsheet Bosonic Correlators}

As usual, the amplitude~\eqref{namp} factorizes into bosonic and fermionic
amplitudes.
The bosonic amplitudes are evaluated using the Green functions
\begin{gather} \label{GreenXl}
\begin{split}
\vev{ X^{\mu} (z, \bar{z})\; X^{\nu} (w, \bar{w})}
= - \apr &\biggl[ g^{\mu \nu} (\log\abs{z-w} - \log\abs{z- \bar{w}})
+ G^{\mu \nu} \log \abs{z - \bar{w}}^2 
\\ &
+ \frac{1}{2 \pi \apr} \theta^{\mu \nu} \log {\frac{z - \bar{w}}{\bar{z}-w}}
\biggr]
\end{split} \\ \label{GreenXt}
\vev{ X^{i} (z, \bar{z})\; X^{j} (w, \bar{w})}
= - \apr g^{i j} (\log\abs{z-w} - \log \abs{z- \bar{w}}).
\end{gather}
Using equations~\eqref{GreenXl} and~\eqref{GreenXt}, we can work
out some basic correlation functions.  For example, with real $y_a$,
\begin{equation} \label{corr1} \raisetag{3\baselineskip}
\begin{split}
A_n  & =  \vev{e^{i q \cdot X} (i) 
\prod_{a=1}^{n} e^{i k_a \cdot X} (y_a)} \\
& =  i \, 2^{\apr k^2}
(2 \pi)^d  \delta(\sum_a k_a + q_{\parallel})
\prod_{a< b}
\abs{\sin (\pi \tau_{ab})}^{2 \apr k_a \cdot k_b} \exp\left[\frac{i}{2} 
\bigl(k_a \times k_b\bigr) \bigl(2 \tau_{ab} -
\epsilon(\tau_{ab})\bigr)
\right] 
\end{split}
\end{equation}
For later convenience we have
expressed
the above correlator in terms of $\tau_a$ which are defined by 
$y_a = - \cot (\pi \tau_a)$. Note that $0 \leq \tau_a \leq 1$ 
follows from $-\infty<y_a<\infty$.
When not specified otherwise, 
the dot product is with respect  to the open string metric and the 
cross product  denotes contraction using $\theta^{\mu\nu}$, i.e. 
$a \times k = a_{\mu} \theta^{\mu \nu} k_{\nu}$.

Another useful correlation function is
\begin{equation} \label{corr2}
B_n  = a_M   \vev{e^{i q \cdot X} (i) \;
 i \dot{X}^M  (y_a) \prod_{b=1}^{n} e^{i k_b \cdot X} (y_b)} 
=  a_M V^M (y_a) A_n 
\end{equation}
with
\begin{multline} \label{bn}
a_M V^M (y_a) = 2 \apr \left[- (a \cdot k)\frac{y_a}{1+y_a^2}
+ \frac{i}{2 \pi \apr} (q \times a) \frac{1}{1+y_a^2}
+ i a_i g^{ij} q_{\perp j} \frac{1}{1+y_a^2} 
\right. \\ \left. + 
\sum_{b=1,b\neq a}^{n} (a \cdot k_b)\frac{1}{y_a-y_b} \right]
\end{multline}
where $k = \sum_b k_b$.
Since we are using pointsplitting
regularization on the worldsheet~\cite{sw}, there are no
$\delta$-functions in~\eqref{bn}.

One more bosonic correlation function that we will need is
\begin{equation} \label{corr3}
\begin{split}
C_n  &= a_{1M} a_{2N}
\vev{e^{i q \cdot X} (i) \;
 i \dot{X}^M  (y_a) \; i \dot{X}^N (y_b) 
\prod_{c=1}^{n} e^{i k_c \cdot X} (y_c)} 
\\ &
=  \left[a_{1M} V^M(y_a) a_{2N} V^N(y_b) 
         + \frac{2\apr}{(y_a-y_b)^2} a_1\cdot a_2\right] A_n.
\end{split}
\end{equation}
Note that $a_1\cdot a_2 = G^{\mu \nu} a_{1\mu} a_{2 \nu} + 
G^{ij} a_{1i} a_{2 j} =  G^{\mu \nu} a_{1\mu} a_{2 \nu} + 
g^{ij} a_{1i} a_{2 j}$.

\subsection{Basic Worldsheet Fermionic Correlators}

The fermionic correlators are somewhat more complicated.  
It is well-known that
\begin{gather} \label{corrt2}
\vev{\psi^{M}(z) \psi^{N} (w)} = g^{MN} \frac{1}{z-w}, \qquad
\vev{\Psi^{M}(y_1) \Psi^{N} (y_2)} = G^{MN} \frac{1}{y_1-y_2}, \\
\vev{\Theta_A(z_1)\Theta_B(z_2)}
= \frac{\ch^{-1}_{AB}}{z_{12}^{5/4}}, \\
\intertext{and}
\label{corrpt2}
\vev{\psi^\mu(z_1) \Theta_A(z_2) \Theta_B(z_3)}
= \frac{\left(\Gamma^\mu \ch^{-1}\right)_{AB}}{\sqrt{2} z_{12}^{1/2}
z_{13}^{1/2} z_{23}^{3/4}}.
\end{gather}
A general correlation function involving an
arbitrary number of $\psi$s and two $\Theta$s is obtained
using the following Wick-like rule. Since the $\psi$s are
like $\Gamma$-matrices (up to normalization), one has
\begin{multline} \label{gencorrt2}
\vev{\psi^{\mu_1}(y_1) \dots \psi^{\mu_n}(y_n)
\Theta_A(z) \Theta_B(\bar{z})}
= \frac{1}{2^{n/2}} \frac{(z-\bar{z})^{n/2-5/4}}%
     {\abs{y_1-z}\dots\abs{y_n-z}} \biggl\{
\left(\Gamma^{\mu_n\dots \mu_1} \ch^{-1}\right)_{AB} \\
+ \; \wick{\psi^{\mu_1}(y_1)\psi^{\mu_2}(y_2)} \;
\left(\Gamma^{\mu_n\dots \mu_3} \ch^{-1}\right)_{AB} \pm \text{perms} \\
+ \; \wick{\psi^{\mu_1}(y_1)\psi^{\mu_2}(y_2)} \;
\wick{\psi^{\mu_3}(y_3)\psi^{\mu_4}(y_4)} \;
\left(\Gamma^{\mu_n\dots \mu_5} \ch^{-1}\right)_{AB} \pm \text{perms} \\
+ \dots \biggr\}.
\end{multline}
The Wick-like contraction, for $y_a$ real, is
\begin{equation} \label{wick}
\wick{\psi^\mu(y_1) \psi^\nu(y_2)} = 
2g^{\mu\nu} 
  \frac{\real\left[(y_1-z)(y_2-\bar{z})\right]}{(y_1-y_2)(z-\bar{z})},
\end{equation}
and in~\eqref{gencorrt2}, the sum is over all possible contractions.
The functional dependence in~\eqref{gencorrt2} and~\eqref{wick}
follows essentially from the conformal weights, and symmetry.
The counterparts of~\eqref{corrpt2}--\eqref{wick} for $\Psi$ can be obtained 
using~\eqref{defPsi}.
Incidentally, note that the fractional power in~\eqref{gencorrt2} will be 
removed by the ghost amplitude.

From equations~\eqref{taft2}
and~\eqref{mforrl}, the fermionic part of~\eqref{namp} 
has the form
\begin{equation} \label{fermic}
\Xi_m =
\vev{\Theta_A(i) \Bigl(\ch \frac{\one-\Gamma^{11}}{2} \fs{C}\Bigr)_{AB}  
\tilde{\Theta}_B(-i) \Psi^{M_1} (y_1)  \cdots \Psi^{M_{2m}} (y_{2m})}
\end{equation}
with various values of $m$. Using~\eqref{mforrl}, the above equation
can be written
\begin{equation}
\Xi_m = T_{AB} \; 
\vev{\Theta_{A}(i) \Theta_{B}(- i) 
\Psi^{M_1} (y_1)  \cdots \Psi^{M_{2m}} (y_{2m}) }
\end{equation}
with
\begin{equation} \label{deftt}
T_{AB} =
\frac{\sqrt{-\det g}}{\sqrt{-\det(g+B)}}
\left( \ch \frac{\one-\Gamma^{11}}{2}
\fs{C} \AE\left(\fs{B}\right) \Gamma^0 \cdots
\Gamma^{p} \begin{cases} \one, \quad& \text{type IIA} \\
                          \Gamma^{11}, & \text{type IIB} \end{cases}
\right)_{AB}.
\end{equation}
In appendix~\ref{app:C*C} we explain how the BRST invariance of the vertex
operator~\eqref{vclose}, implies that $C$ is in a special gauge in which
$C$ and $F$ are both self dual: 
$\Gamma^{11} \fs{C} = -\fs{C}$ and $\Gamma^{11} \fs{F} = \Gamma^{11}
\fs{dC} = -\fs{F}$. Thus equation~\eqref{deftt} 
yields an identical form for both the IIA and IIB theories
\begin{equation} \label{deft}
T_{AB} =
\frac{\sqrt{-\det g}}{\sqrt{-\det(g+B)}}
\left(\ch \fs{C} \AE\left(\fs{B}\right) \Gamma^0 \cdots
\Gamma^{p} \right)_{AB}.
\end{equation}

Using equation~\eqref{defPsi} to turn the correlation
function~\eqref{gencorrt2}
involving left-moving fermions into a correlation function for boundary
fermions, it can be seen
that~\eqref{fermic} and eventually the amplitude~\eqref{namp} can be 
written as a sum 
\begin{equation}
{\mathcal A}_n  =  \sum_{m=0}^n \omega_{M_1 \cdots M_{2m}}  
\Lambda^{M_1 \cdots M_{2m}},
\end{equation}
where $w_{\cdots}$ contain all the dependence on the polarizations
and momenta of the open string modes, and  $\Lambda^{\cdots}$ are
the Gamma matrix traces
\begin{smaleq} 
\begin{equation} \label{defL}
\begin{split} \raisetag{1.25\baselineskip}
\Lambda^{M_1\dots M_{2m}} 
&= \left(\frac{1}{g-B}g\right)^{M_1}{_{N_1}} \dots
   \left(\frac{1}{g-B}g\right)^{M_{2m}}{_{N_{2m}}}
\Tr\left[\left(\Gamma^{N_{2m}\dots N_{1}} \ch^{-1}\right)^\transpose
   \ch \fs{C} \AE(\fs{B}) 
   \Gamma^0 \cdots \Gamma^p\right].
\\ &
 = -\left(\frac{1}{g-B}g\right)^{M_1}{_{N_1}} \dots
   \left(\frac{1}{g-B}g\right)^{M_{2m}}{_{N_{2m}}}
\Tr\left(\Gamma^{N_1\dots N_{2m}} \fs{C} \AE(\fs{B}) 
   \Gamma^0 \cdots \Gamma^p\right).
\end{split}
\end{equation}
\end{smaleq}%
Note that the $\Tr$ above is taken only with respect to the spacetime 
spinor indices and that we have used the property of the 
charge conjugation matrix $\ch$,
\begin{equation}
\left(\Gamma^{N_{2m}\dots N_{1}} \ch^{-1}\right)^\transpose
= - \Gamma^{N_1\dots N_{2m}} \ch^{-1} \ .
\end{equation}

By evaluating the trace in~\eqref{defL} explicitly, in 
appendix~\ref{app:trace} we show that (up to a sign that
depends only on $p$)
\newif\ifexpgaveL\expgaveLfalse%
\newcommand\expgiveL{%
\begin{equation} \ifexpgaveL\tag{\ref{expgiveL}}\else\label{expgiveL}\fi
\frac{\omega_{\mu_1\dots\mu_q}}{q!}\frac{\chi_{i_1\dots i_r}}{r!}
   \Lambda^{\mu_1\dots \mu_q i_1\dots i_r}
= 32 \left\{
\star \left[(e^{-\co{\sfrac{\theta}{2\pi\apr}}} \omega) \, 
 (\co{\chi} C) e^B\right]
\right\}_{\text{0-form}}
\end{equation}}%
\newcommand\putexpgiveL{\expgiveL\expgaveLtrue}
\putexpgiveL
where wedge products
are implied and the subscript means that we keep only the scalar part of the
right-hand side.  The Hodge dual, $\star$, is
the Hodge dual in the {\em worldvolume\/} of the D-brane, with
respect to the {\em closed\/} string metric.
The notation $\co{T}$ denotes contraction with respect to
the antisymmetric tensor $T$ of rank $m$; i.e.\
\begin{equation} \label{defco}
\left(\co{T} C^{(n)}\right)_{M_1 \dots M_{n-m}}
= \frac{1}{m!} T^{N_m \cdots N_1 } C_{N_1 \cdots N_m M_1 
\dots M_{n-m}}.
\end{equation}
In~\eqref{expgiveL}, $\chi$ only contracts with the transverse indices 
of $C$ and 
\begin{equation}
e^{- \co{\theta}} \omega 
= \sum_{n} \frac{(-1)^n}{n!} 
\overbrace{\co{\theta} \cdots \co{\theta}}^{\text{$n$ copies}} \omega .
\end{equation}
Note that in equation~\eqref{expgiveL}, $\co{\theta}$ does not
contract with objects outside the
parentheses.

\section{The Amplitudes} \label{sec:amp}

In this section we will compute the amplitudes with one and two open
strings explicitly.  The contribution to~\eqref{namp} from 
ghosts is always
\begin{equation} \label{ghost}
{\cal A}_{ghost} = \vev{c(i) \tilde{c}(-i) c(0)} \vev{e^{-\phi/2} (i) \,
e^{-3 \tilde{\phi}/2} (-i)} = (2i)^{1/4}.
\end{equation}
where we have used 
\begin{equation}
\vev{e^{-\phi/2}(z) e^{-3\tilde{\phi}/2}(\bar{z}) }
= \frac{1}{(z-\bar{z})^{3/4}}, \qquad
\vev{c(z_1) \tilde c(\bar z_2) c(z_3)} = (z_1 - \bar z_2) 
(\bar z_{2} -z_3) (z_1 - z_{3}).
\end{equation}

\subsection{One Open String} \label{sec:one}

We wish to evaluate
\begin{smaleq}
\begin{equation}
{\mathcal A}_1 = (2 i)^{1/4} \lambda g_c \, 
a_M \vev{\Theta(i) \ch \frac{\one-\Gamma^{11}}{2} \fs{C} \tilde{\Theta}(-i) 
e^{i q \cdot X(i)}
\left(i \dot{X}^{M}(0)
+ 2\apr k \cdot \Psi(0) \Psi^{M}(0) \right) e^{i k \cdot X(0)} },
\end{equation}
\end{smaleq}%
where the factor $(2 i)^{1/4}$ comes from~\eqref{ghost}.
Using the rules of the previous section, it is straightforward to
evaluate this.  One finds,
\begin{align} \label{founda1} \raisetag{\baselineskip}
\begin{split}
{\mathcal A}_1 & = \frac{1}{2 \pi} \lambda g_c 
C_{D_2} \frac{\sqrt{-\det g}}{\sqrt{-\det(g+B)}}
(2\pi)^{p+1} \delta^{(p+1)}(k+q_\parallel)
\left[
i (2 \pi \apr) k_\mu a_M \, \Lambda^{\mu M}
+ \cM \Lambda \right], \\
& =  \frac{\lambda}{2} \kappa_{10} \, \mu_{p}  
\sqrt{-\det g} \, (2\pi)^{p+1} \delta^{(p+1)}(k+q_\parallel)
\left[ i (2 \pi \apr) k_\mu a_M \, \Lambda^{\mu M}
+ \cM \Lambda \right],
\end{split}
\end{align}
with  
\begin{equation} \label{wilk}
\cM = i q \times a  + i (2 \pi \apr) a \cdot q_{\perp}.
\end{equation}
$\Lambda^{\mu M}$ and $\Lambda$ were defined in equation~\eqref{defL}
and in the second line we have used 
equations~\eqref{defym}--\eqref{tension} to obtain
\begin{equation}
g_{c} \, C_{D_2} \frac{\sqrt{-\det g}}{\sqrt{-\det (g+B)}}
= \pi \kappa_{10} \, \mu_{p} \, \sqrt{-\det g},
\end{equation}
where $\mu_{p} = T_p =  \frac{1}{(2 \pi)^{p} g_s \apr^{\frac{p+1}{2}}}$
is the \RR\ charge density of a D$p$-brane.

Of course,
the entire amplitude~\eqref{founda1} consists of contact
terms that will give us the field theory action.  
This will not be true of the amplitude with two open strings,
for which there are also
poles corresponding to intermediate states.  In
particular, the two open string amplitude will have 
$\apr$ corrections, while
equation~\eqref{founda1}
is exact in $\apr$.

\subsection{Two Open Strings} \label{sec:two}

Now we wish to evaluate
\begin{smaleq}
\begin{multline} \label{twoamp}
{\mathcal A}_2 = (2i)^{1/4} \lambda g_c
\int_{-\infty}^\infty dy
\left\langle\Theta(i) \ch  \frac{\one-\Gamma^{11}}{2} \fs{C}
\tilde{\Theta}(-i)
e^{i q \cdot X(i)}
\right. \\ \times \left.
\left(i \dot{X}^{M}(0)
+ 2\apr k_1 \cdot \Psi(0) \Psi^{M}(0) \right) e^{i k_1 \cdot X(0)}
\left(i \dot{X}^{N}(y)
+ 2\apr k_2 \cdot \Psi(y) \Psi^{N}(y) \right) e^{i k_2 \cdot X(y)}
\right\rangle
a_{1M} a_{2N}  \\ 
 = \lambda \pi \kappa_{10} \mu_p \sqrt{-\det g}
\int_{-\infty}^\infty dy \; \left[I_0 (y) + I_2 (y) + I_4 (y) \right]
\end{multline}
\end{smaleq}%
where $I_{0,2,4}$ correspond to the correlators involving
respectively zero, two and 
four $\Psi$s, and we have extracted from
them an appropriate  prefactor.

Using the formulas of the previous section, it is straightforward to
find
\begin{equation} \label{i00}
\begin{split}
I_0 & = - \frac{i}{2} C_2 (y) \, \Lambda  \\
& = - \frac{i}{2} A_2 (y) \, \Lambda 
\left[ 2 \apr (a_1 \cdot a_2) \frac{1}{y^2} 
+ \frac{1}{\pi^2} \cM_1 \cM_2 \frac{1}{1+ y^2} 
- (2 \apr)^2 (a_1 \cdot k_2) (a_2 \cdot k_1)  \frac{1}{y^2 (1+ y^2)} \right.
\\
& \qquad \qquad \left.
-\frac{2 \apr}{\pi} \bigl[(a_1 \cdot k_2) \cM_2 - (a_2 \cdot k_1) \cM_1\bigr]
\frac{1}{y (1+ y^2)}  \right]
\end{split}
\end{equation}
where $C_2$, $A_2$ and $\Lambda$ were given in
equations~\eqref{corr3}, \eqref{corr1}
and~\eqref{defL} respectively and $\cM_a$ was defined in~\eqref{wilk}
with $a$ labeling the gauge boson. Similarly,
\begin{equation} \label{i2}
I_2 = \frac{\apr} { \pi}  \frac{A_2 (y) }{1+ y^2} (Q_1 \cM_2 + Q_2 \cM_1)
- 2 \apr^2 \frac{A_2 (y) }{y (1+y^2)} 
\left[(a_1 \cdot k_2) Q_2 - (a_2 \cdot k_1) Q_1 \right]
\end{equation}
where we have defined a short-hand notation
\begin{equation}
Q_a =  k_{a \mu} a_{a N} \Lambda^{\mu N}.
\end{equation}

The expression for $I_4$ is more complicated. According to~\eqref{gencorrt2}
it is convenient to split it into three parts 
$I_{4} = I_{40} + I_{42} + I_{44}$ 
where the second index denotes the number of $\Gamma$-matrices appearing 
in~\eqref{gencorrt2}. We find that 
\begin{align} \label{I40}
I_{40} & 
= - 2 i \apr^2 \frac{A_2(y)}{y^2 (1+y^2)} \Lambda 
\left[(a_1 \cdot k_2) (a_2 \cdot k_1) - 
      (k_1 \cdot k_2) (a_1 \cdot a_2)\right],
\\
\begin{split} \label{I42}
I_{42} =& -2 \apr^2 \frac{A_2(y)}{y (1+y^2)}
\left[(a_1 \cdot k_2) k_{1 \mu} a_{2 N} \Lambda^{\mu N} 
- (a_1 \cdot a_2) k_{1 \mu} k_{2 \nu} \Lambda^{\mu \nu} \right.\\ 
& \left.  - (a_2 \cdot k_1)  k_{2 \mu} a_{1 N} \Lambda^{\mu N}
- (k_1 \cdot k_2) a_{1 M} a_{2 N} \Lambda^{MN} \right],
\end{split}
\\ \label{I44}
I_{44} & = 2 i \apr^2 \frac{A_2(y)}{1+ y^2} k_{1 \mu} a_{1 M}
 k_{2\nu} a_{2 N}
 \Lambda^{\mu M \nu N}.
\end{align}
Since both $I_0$ and $I_{40}$ are proportional to $\Lambda$ we may combine
them to obtain,
\begin{equation}
\begin{split} \label{i0}
I_0 + I_{40} & = - A_2 (y) \Lambda \left[
\frac{i}{2 \pi^2} \cM_1 \cM_2 \frac{1}{1+ y^2} 
- \frac{i \apr}{\pi} \bigl[(a_1 \cdot k_2) \cM_2 - (a_2 \cdot k_1) \cM_1\bigr]
\frac{1}{y (1+ y^2)} \right.\\
& \left.  + i \apr (a_1 \cdot a_2) 
\left((1 + \apr t) \frac{1}{y^2} - \apr t \frac{1}{1+y^2}\right)
\right],
\end{split}
\end{equation}
with $t = - 2 k_1 \cdot k_2$.

We now proceed to evaluate the $y$-integrals in~\eqref{twoamp}, which can
be found in Gradshteyn and Ryzhik~\cite{gr}
equations~(3.631.9) and~(3.633.1). We find%
\footnote{We give the complete expressions in appendix~\ref{sec:ints}.}
\begin{subequations} \label{ints}
\begin{align} \label{int0}
& \int_{-\infty}^\infty dy \; \frac{A_2 (y)}{1+ y^2}  =
\pi i \frac{\sin \frac{k_1\times k_2}{2}}{\frac{k_1\times k_2}{2}} 
+ \order{\apr t},
\\ \label{intcot2}
& \int_{-\infty}^\infty dy \; \frac{A_2 (y)}{y (1+ y^2)}  =
 \frac{2}{\apr t} \sin\frac{k_1\times k_2}{2}
+\order{\apr t},
\\ \label{intcsc2}
& (1 + \apr t) \int_{-\infty}^\infty dy \; \frac{A_2 (y)}{y^2}  =
\pi i \apr t \frac{\sin \frac{k_1\times k_2}{2}}{\frac{k_1\times k_2}{2}}
- \frac{2i}{\pi} \frac{k_1\times k_2}{\apr t}  \sin\frac{k_1\times k_2}{2}
+ \order{\apr t},
\end{align}
\end{subequations}
where $A_2 (y)$ is given by~\eqref{corr1} with $n=2$ and $y_1 =0$. 
On the right-hand side of equation~\eqref{ints} we have, for
notational simplicity, suppressed
the factor \hbox{$(2\pi)^{p+1} \delta(k_1+k_2+q_\parallel)$}.

We see from~\eqref{ints} that there are $\apr$ corrections to the
amplitudes.  We shall be interested only in the lowest order 
contact terms in~\eqref{twoamp}. The terms containing poles 
in $\apr t$  may be understood from processes involving 
exchanging  intermediate Yang-Mills modes using vertices~\eqref{founda1}
and those of noncommutative Yang-Mills theory. 
Substituting~\eqref{ints} into~\eqref{i2}--\eqref{i0} and collecting 
only the lowest order contact terms we find
\begin{multline} \label{founda2}
{\mathcal A}_{2}  = \frac{\lambda}{2} \kappa_{10} \mu_p \sqrt{-\det g} \,
\biggl\{ 
\frac{\sin \frac{k_1\times k_2}{2}}{\frac{k_1\times k_2}{2}} 
\Bigl[\cM_1 \cM_2 \Lambda 
- (2 \pi \apr)^2 
k_{1 \mu} a_{1 M} k_{2 \nu} a_{2 N}  \Lambda^{\mu M \nu N} 
\\
+ i (2 \pi \apr) \bigl(k_{1 \mu} a_{1 N} \cM_2
   + k_{2 \mu} a_{2 N} \cM_1\bigr) \Lambda^{\mu N} 
\Bigr]
- 4 \pi \apr \sin \frac{k_1\times k_2}{2} a_{1M} a_{2N} \Lambda^{MN} 
\biggr\}
\end{multline}
where we again have suppressed 
the factor \hbox{$(2\pi)^{p+1} \delta(k_1+k_2+q_\parallel)$}.

\subsection{The Total Amplitude} \label{sec:totamp}

The last term of~\eqref{founda2} is proportional to
$\sin \frac{k_1 \times k_2}{2}$ and it precisely combines with the first 
term in~\eqref{founda1} to give
\begin{equation} \label{ff}
\frac{\lambda}{2} \kappa_{10} \mu_p \sqrt{-\det g} \, 
(\pi \apr) f_{MN} \Lambda^{MN},
\end{equation}
with 
\begin{subequations} \label{split}
\begin{align} \label{fncf}
f_{\mu \nu}&  = i (k_{\mu} a_{\nu} - k_{\nu} a_{\mu}) 
- i \com{a_{\mu}}{a_{\nu}}_{\ast},
\\ \label{covd}
f_{\mu i} & = - f_{i \mu}  = D_{\mu} a_i = i k_\mu a_i 
               - i \com{a_\mu}{a_i}_{\ast},
\\
f_{ij} & = - i \com{a_i}{a_j}_{\ast},
\end{align}
\end{subequations}
where in momentum space 
\begin{equation} 
\com{f(k_1)}{g(k_2)}_{\ast} = 
- 2 i \sin \frac{k_1 \times k_2}{2}
f(k_1) g(k_2) 
\end{equation}

We recognize the factor
$\frac{\sin \frac{k_1\times k_2}{2}}{\frac{k_1\times k_2}{2}}$
in~\eqref{founda2} 
as the $\ast_2$-operation~\cite{garousi,lm2}, which in momentum
space is
\begin{equation} \label{ast2}
f(k_1) \ast_2 g(k_2) = 
f(k_1) \frac{\sin \frac{k_1 \times k_2}{2}}{\frac{k_1 \times k_2}{2}} g(k_2) 
\end{equation}
We  will now combine the two amplitudes~\eqref{founda1} 
and~\eqref{founda2}. Using~\eqref{split} and~\eqref{ast2} 
we find that  ${\mathcal A}_1+{\mathcal A}_2$ corresponds to the
action
\begin{multline} \label{totamp}
S = \frac{\lambda}{2} \kappa_{10} \mu_{p} \int \sqrt{-\det g}
\left\{
(1 + \cM +  \half \cM \ast_2 \cM) \Lambda 
\right. \\ \left.
+ \pi \apr \left(f_{MN} +  \cM \ast_2 f_{MN} \right) \Lambda^{MN}
+ \half  (\pi  \apr)^2 f_{MN} \ast_2 f_{PQ} \Lambda^{MNPQ}
\right\} .
\end{multline}
Equation~\eqref{totamp} contains some terms that
are cubic and quartic in open
string modes which therefore do not follow directly from the amplitudes we
computed.  But, of course, we expect these terms by gauge invariance.
Also, we have inserted a $\Lambda$ by hand, corresponding to a tadpole
diagram with just the \RR\ field.

We recognize the factors of $\cM$ appearing in~\eqref{totamp} precisely
as expected from the expansion of a straight open Wilson line  
(with the substitution \hbox{$a_\mu \rightarrow
A_{\mu}$}, \hbox{$2 \pi \apr a_i \rightarrow X^i$})
\begin{equation} \label{cwilson} 
W(x,\cC_q)  =  P_{\ast} \exp \left[ i 
\int_0^1 d \tau  \,
\left(q_{\mu} \theta^{\mu \nu} A_{\nu} (x + \xi (\tau))
+ q_{\perp i}  X^i ( x + \xi (\tau)) \right) \right]
\end{equation}
with the path
$\xi: \cC_q \hookrightarrow {\mathbb{R}}^{p+1}$ given by
$\xi^{\mu} (\tau) = \theta^{\mu \nu} q_{\nu} \tau$.
We also see the $n$-ary operations
(in this case $\ast_2$) appearing just as we expect from smearing 
the Yang-Mills operators along a Wilson line~\cite{liu}. 

\section{The Full WZ Term} \label{sec:wz}

We now attempt to extract the full \WZ\ coupling from the first
few terms given in~\eqref{totamp}. For simplicity,
we first focus on the ``zero-momentum''
couplings (i.e. setting $q$, the momentum of the \RR\ potential, to zero, 
in which case $\cM=0$),
\begin{equation} \label{amp0}
S =  \frac{\lambda}{2} \kappa_{10} \mu_{p} \int \sqrt{-\det g} 
\Tr \left[ \Lambda
+ 2\pi \apr \frac{f_{MN}}{2!} \Lambda^{MN}
+ \frac{1}{2} (2\pi \apr)^2 \frac{f_{MN}}{2!} \frac{f_{PQ}}{2!}
  \Lambda^{MNPQ}
\right].
\end{equation}
Although in the last section we have not included the Chan-Paton factors
explicitly, they can be added in straightforwardly. 
For this reason we have included a $\Tr$ over the $U(n)$ indices
in~\eqref{amp0} for the case of $n$ D-branes.
Using  equation~\eqref{expgiveL} for the 
$\Lambda^{\cdots}$s, we find that~\eqref{amp0} can be precisely 
reproduced by expanding the formula 
\begin{equation} \label{rrf}
S =
\mu_{p} \STr \int  \left(e^{-\co{\sfrac{\theta}{2\pi\apr}}} 
e^{2\pi \apr f} 
      \Pb\right)
  e^{-2\pi \apr i  \co{\com{\phi}{\phi}}} \,  C  e^B,
\end{equation}
to terms quadratic in open string modes. 
In reaching~\eqref{rrf} we have 
set the value of the normalization constant $\lambda = 1/16$, and 
for notational convenience we have made the
substitution $a_i \rightarrow \phi_i$
and absorbed the factor of $\kappa_{10}$ into $C$.   
In~\eqref{rrf} when evaluating products of open-string fields, the
$\ast$-product is implied and $\STr$ is the symmetrized trace over 
both the $U(N)$ matrices and $\ast$-product.
As usual wedge products are implied in expanding the 
exponentials and in product of forms, and the integration 
extracts only the $(p+1)$-form in the integrand. $\co{T}$ denotes 
contraction with respect to an antisymmetric tensor as in the
definition~\eqref{defco}. The notation $\Pb$ denotes the pullback;
e.g.\ 
\begin{equation} \label{defpb}
\Pb \omega^{(2)}_{\mu\nu} = \omega^{(2)}_{\mu\nu}
+ 2 (D_{[\nu} X^i) \omega^{(2)}_{\mu]i}
+ D_{\mu} X^i D_{\nu} X^j \omega^{(2)}_{ij}.
\end{equation}
where $D_{\mu} X^i = 2\pi\apr D_{\mu} \phi^i$ should be understood as
the coordinate space
version of~\eqref{covd}.  
The parenthesis in~\eqref{rrf} enforce that $\theta$ can contract with
the longitudinal indices coming from the pullback, but not with $C$ or $B$.
For example, with $C^{(2)} = \left[\frac{1}{2}C_{\mu\nu} dx^{\mu} dx^{\nu}
      + C_{\mu i} dx^\mu dx^i
      + \frac{1}{2} C_{ij} dx^i dx^j \right]$,
\begin{multline} \label{ex(iP)C}
\left(e^{-\co{\theta}} \Pb\right) C^{(2)}
= \left[\frac{1}{2}C_{\mu\nu} dx^{\mu} dx^{\nu}
      + C_{\mu i} dx^\mu dx^i
      + \frac{1}{2} C_{ij} dx^i dx^j \right]
\\
+ D_{\mu} X^i \left[C_{i\nu} dx^\mu dx^\nu + C_{ij} dx^\mu dx^j
      \right]
+ D_{\mu} X^i D_{\nu} X^j C_{ij} dx^\mu dx^\nu
- \frac{1}{2} \theta^{\tau\sigma} D_{\sigma} X^i D_{\tau} X^j C_{ij}.
\end{multline}

An explicit expansion 
of~\eqref{rrf} and its comparison with~\eqref{amp0} is given in 
appendix~\ref{sec:verify}. Since there are a rather large number of terms
involved,
we see this as very strong evidence that~\eqref{rrf} gives 
the correct couplings involving an arbitrary number of open string modes.
Note that at quadratic order in open string modes, $\STr$ in~\eqref{rrf}
reduces to the normal trace and the $\ast$-product reduces to the ordinary 
product as in~\eqref{amp0}; thus at this order we have only checked 
the tensor structure and not the product and ordering structure.  
In section~\ref{sec:wepb} we give further evidence for
equation~\eqref{rrf}
by looking at 
amplitudes with an arbitrary number of open string modes.

For couplings to \RR\ potentials with nonzero momentum $q$, 
as already discussed around equation~\eqref{cwilson}, 
we include a straight open  Wilson line~\eqref{cwilson} with its 
length given by $\theta^{\mu \nu} q_{\nu}$ and smear the
operators in~\eqref{rrf}  over the Wilson line~\cite{liu,dastrivedi}.  
The full action is therefore,
\begin{smaleq}
\begin{equation} \label{rrfw}
S = \mu_p  \int d^{10} q \biggl\{
\int d^{p+1} x \, L_\ast \left[ W(x,\cC_q)
\left(e^{-\co{\sfrac{\theta}{2\pi\apr}}} 
e^{2\pi \apr f}  \Pb\right)
  e^{-2 \pi \apr i  \co{\com{\phi}{\phi}}} e^{i q \cdot x}\right] 
C(q)  e^B \biggr\},
\end{equation}
\end{smaleq}%
where $L_\ast$ denotes the prescription of smearing of all operators in 
the integrand over the Wilson line with the path ordering in terms of 
the $\ast$-product. For this purpose, 
$f_{\mu \nu}, D_{\mu} \phi$ and $\com{\phi}{\phi}$ are considered 
individual  operators.
For example,
\begin{align} \label{giop}
\begin{split} \raisetag{3\baselineskip}
&  \int d^{p+1} x \; L_{\ast} \left[ W(x,\cC_q) f_{\mu \nu}(x) 
f_{\lambda \rho} (x) \right] \ast e^{iq_{\mu} x^{\mu}} \\
& =   \int d^{p+1} x  \!
 \int^{1}_{0}\! d \tau_1 d \tau_2
 \; P_{\ast} \left[ W(x,\cC_q) 
     f_{\mu \nu} (x + \xi(\tau_1))
f_{\lambda \rho} (x + \xi(\tau_2))
 \right] \ast e^{i q _{\mu} x^{\mu}} \\
& = \sum_{n=0}^\infty \frac{i^n}{n!} \int d^{p+1} x 
\int_0^1 d\tau_1 \cdots \int_0^1  d \tau_{n+2} \;
P_\ast \biggl[ f_{\mu \nu} (x + \xi(\tau_1)) \ast 
f_{\lambda \rho} (x + \xi(\tau_2)) \ast  \biggr. \\
& \biggl. \qquad
{\mathcal M} (x + \xi(\tau_3)) \ast \cdots \ast 
{\mathcal M} (x + \xi(\tau_{n+2})) \biggr] \ast e^{i q \cdot x}
\end{split}
\end{align}
where ${\mathcal M}= i q \times a + i(2 \pi \apr) q_{\perp} \cdot \phi$ comes 
from the expansion of the Wilson line~\eqref{cwilson} and 
$P_{\ast}$ denotes path ordering.  
On performing the $\tau$ integrations, equation~\eqref{giop}
can be written in terms of a power series in $\cM$ using 
$n$-ary operations $\ast_n$~\cite{garousi,lm2}.
These factors of ${\mathcal M}$ were seen in equation~\eqref{totamp}.
In appendix~\ref{sec:verify} we verify also that the quadratic
expansion of~\eqref{rrfw} gives precisely~\eqref{totamp}.
The definition and properties 
of the $\ast_n$ $n$-ary operations, and their the relations to 
the expansion of open Wilson lines, was given in~\cite{liu}.
Note that it is manifest from the above equations
that the $L_{\ast}$-prescription 
completely symmetrizes the integrand. Again, in section~\ref{sec:wepb}
we shall give further evidence for equation~\eqref{rrfw} by looking 
at higher order terms.

We conclude this section with some remarks:
\begin{enumerate}

\item As we take $\theta \rightarrow 0$, the Wilson line collapses to a point 
and we recover the standard commutative result, including the 
factor $e^{-2 \pi \apr i  \co{\com{\phi}{\phi}}}$ 
found in~\cite{myers,taylor}.

\item As we take the
zero momentum limit, $q  \rightarrow 0$, the Wilson 
line collapses to a point and the $\ast_n$ operations become 
symmetrized $\ast$-products. More explicitly, 
\begin{equation}
\Tr \int d^{p+1} x \ast_n \left[f_1 (x) \cdots f_n (x) \right] 
= \STr \int d^{p+1} x \; (f_1 (x) \ast \cdots \ast f_n (x))
\end{equation}
where on the right  hand of the equation $\STr$ denotes the symmetrized 
trace prescription denoting the normalized sum over all
possible permutations.

\item When we take $q=0$ and set the transverse scalar fields to zero,
\eqref{rrfw} becomes
\begin{equation} \label{grr}
\swz  = \mu_{p} \, 
\Tr_{\theta}  \left( e^{- \co{\theta}} e^{f} \right)\; e^{B} C 
\end{equation}
where for convenience we have set $2 \pi \apr =1$ and defined 
$\Tr_{\theta} = \Tr \int$.  The combination of $C e^B$ is a consequence  
of T-duality.\cite{bmz,fot}\ 
Equation~\eqref{grr} implies that the \RR\ charges 
of the lower dimensional D-branes generated  by the topologically 
nontrivial configurations of the gauge theory are given by
\begin{equation} \label{ell}
\Tr_{\theta}  \left( e^{- \co{\theta}} e^{f} \right)
\end{equation}
which is called the Elliott formula in noncommutative geometry.
This agrees very well with the noncommutative geometry result that
Elliott's formula is integer valued, while the Chern characters are not.
That~\eqref{ell} gives the right charges for D-branes was 
anticipated in refs.~\cite{schwarz,schwarzre} based on the 
K-theory of the noncommutative torus. For more discussion of 
equations~\eqref{grr} and~\eqref{ell}, see section~3 of~\cite{lm4}.

\item The result~\eqref{rrfw} was motivated from the on-shell amplitudes
and corresponds to the $\Phi=0, \theta = - (2 \pi \apr) \frac{1}{g+B} B 
\frac{1}{g-B}$ description of the D-brane couplings. 
In~\cite{lm4} we argued that it actually applies to every
$\theta$-descriptions (see also the next section).

\end{enumerate} 

\section{Comparison to the Matrix Model} \label{sec:mm}

In this section we shall rewrite the couplings~\eqref{rrf} 
and~\eqref{rrfw} in various other forms. In particular we shall 
make connections to the results of~\cite{mukhi,ooguri,mukhi2}
which describe the couplings in the $\theta = \frac{1}{B}, \Phi=-B$ 
description. For convenience in this section, we will 
set $2\pi\apr =1$ and
assume that
$\theta$ and $B$ have maximal rank. Then, for $p$ odd (IIB), 
we shall consider a Euclidean world-volume with all longitudinal 
directions noncommutative, while for $p$ even (IIA) all longitudinal 
directions but time are noncommutative.

In~\cite{mukhi,ooguri,mukhi2}, motivated from the connection between 
noncommutative gauge theory and the Matrix 
model (see e.g.~\cite{seiberg}), the zero  momentum \RR\ couplings were  
argued to be
\begin{equation} \label{asX}
\swz = \frac{\mu_{p}}{\Pf(\theta)} \STr 
\int \star \left(e^{-i \co{\com{X}{X}}} \; C \right),
\end{equation}
with, 
\begin{equation} \label{ncb}
X^{\mu} = x^{\mu} + \theta^{\mu \nu} a_{\nu}, \quad
X^{i} = \phi^{i} (x^{\mu}), \quad 
   \com{x^{\mu}}{x^{\nu}} = i \theta^{\mu \nu},
\end{equation}
where now $\star$ is the Hodge dual in the {\em noncommutative\/}
directions of the brane, still with respect to the {\em closed\/}
string metric.
That is, for odd $p$ (IIB), we take the Hodge dual in the brane,
whereas for even $p$, we take the Hodge dual along the spatial
directions of the brane.
Thus, for odd $p$, equation~\eqref{asX} is equivalent to
$\swz = \frac{\mu_p}{\Pf(\theta)} \STr 
\int
d^{(p+1)} x \, \sqrt{g}
\left(e^{-i \co{\com{X}{X}}} \; C \right)$.
For even $p$, equation~\eqref{asX} can be similarly rewritten as the
integral of a time-like one-form.
Equation~\eqref{asX} is essentially the \RR\ coupling of  D-instantons 
expanded around a background~\eqref{ncb} which is noncommutative. 
As noted in~\cite{seiberg}, the Matrix model description corresponds to 
the  $\theta = \frac{1}{B}, \Phi=-B$ description of the D-brane.
Note that 
\begin{equation} \label{defmf}
\com{X^{\mu}}{X^{\nu}} = i \left(\theta - \theta f \theta \right)^{\mu \nu},
\quad 
\com{X^{\mu}}{X^i} = i \theta^{\mu \nu} D_{\nu} \phi^i.
\end{equation}

In the following we shall show that equation~\eqref{rrf} can be written 
in terms of Matrix model-type variables~\eqref{ncb} as 
\begin{equation} \label{asXX}
\swz = \frac{\mu_{p}}{\Pf(\theta)} \STr 
\int  \star
\left(e^{-i \co{\com{X}{X}}} \; 
e^{B-\frac{1}{\theta}} C \right),
\end{equation}
It is remarkable that~\eqref{asX} and~\eqref{asXX} are so tantalizingly 
close to each other in this form%
\footnote{This apparent similarity is actually somewhat deceptive. 
Explicit expansion when $B$ is far from infinite, shows that
they actually differ significantly since the $\mu, \nu$
indices in $\com{X^{\mu}}{X^{\nu}}$ and $\com{X^{\mu}}{X^i}$ also contract 
with $e^{B- \frac{1}{\theta}}$. Thus the on-shell amplitudes which follow 
from~\eqref{asX} and~\eqref{asXX} are very different for generic 
values of $B$, agreeing only in the limit of infinite $B$.}. 
In particular if we take $\theta = \frac{1}{B}$ in~\eqref{asXX} we precisely 
recover~\eqref{asX}. The reason may be understood from two aspects:
\begin{enumerate} 

\item When we take the large $B$ limit, the worldsheet boundary conditions 
reduce to that of D-instantons and $\theta = - \frac{1}{g+B} B 
\frac{1}{g-B} \rightarrow \frac{1}{B}$. Thus we precisely 
recover~\eqref{asX} from~\eqref{asXX} in this limit.
    
\item   In~\cite{lm4} based on the topological nature 
of the terms in~\eqref{asXX}, we argued that~\eqref{asXX} (while derived 
in the $\Phi=0$ description) gives the leading  term (in an $\apr$ expansion) 
for every $\theta$-description. In particular it  applies to the Matrix 
description with $\theta = \frac{1}{B}, \Phi = -B$. We emphasize that 
while different descriptions are related by field redefinitions, they give 
rise to the same on-shell amplitudes only after we sum over all
$\apr$ corrections.

\end{enumerate}

To derive~\eqref{asXX} from~\eqref{rrf} we  introduce a shorthand notation
for the pullback. With a mild notational abuse,
\begin{equation} \label{altPb}
\Pb = e^{D\co{\phi}} = e^{D\co{X}}.
\end{equation}
where $D\co{\phi}= D_{\mu} \phi^i d x^{\mu}$ is considered 
as a one-form in the worldvolume and a contracted vector in the transverse 
dimensions.  That is, we can think of $D\co{\phi}$ as an
operator which acts on forms to the right, by contracting the vector
index and antisymmetrizing the form index, thereby preserving the
dimension of the form on which it acts.  This reproduces~\eqref{defpb}.

Note the following identities
for the manipulations of forms and contractions%
\footnote{We emphasize that the Hodge dual, $\star$, is only in the
(Euclidean) noncommutative directions of the brane.  In particular, the $n$ in
equation~\eqref{itow} is the dimension of the form along the
noncommutative directions. Equations~\eqref{moveco} and~\eqref{intc} 
are consequences
of~\eqref{itow} and $\int (\star \omega) (\star \chi) = \int \omega \chi$
where the integral, here and in the text,
is over the entire worldvolume of the brane.  Finally, we should note
that, in a basis in which $\theta$ is skew-diagonal, 
our convention for
\newcommand\pfdim{\Delta}%
the Pfaffian of a $(2\pfdim)$-dimensional antisymmetric matrix $M$
is
\hbox{$\Pf M = \frac{(-1)^{\pfdim}}{2^{\pfdim} \pfdim!}
  \epsilon_{\mu_1\cdots\mu_{2\pfdim}}
  M^{\mu_1\mu_2} \cdots M^{\mu_{2\pfdim-1} \mu_{2\pfdim}}$}.
That is, to compute $\Pf\theta$, one can go to a basis in which
$\theta$ is skew diagonal, and
multiply over the skew eigenvalues of $\theta$ in the lower-left
corners of the $2\times 2$ blocks.%
}%
\begin{gather}
\label{itow}
e^{-\co{\theta}} \omega^{(n)} = (-1)^{n+\frac{p(p+1)}{2}}
    \Pf(\theta) \star \left( \left(\star
e^{-\theta^{-1}} \right) (\star \omega^{(n)}) \right), \\
\label{moveco}
\int \left(e^{-\co{\theta}} \chi\right) \omega 
= \int (e^{-\co{\theta}} \omega) \chi \\
\label{intc}
\int \star \left(e^{-\co{\theta}} \omega\right) 
= (-1)^{\frac{p(p+1)}{2}} \Pf(\theta) \int e^{-\theta^{-1}} \omega.
\end{gather}
Using
\begin{align}
\com{D\co{\phi}}{\co{\theta}} &=  -\co{\theta D \phi}, &
\com{D\co{\phi}}{\co{\theta D \phi}} &= 2\co{\co{\theta}(D\co{\phi})^2}, &
\com{\co{\theta D\phi}}{\theta^{-1}} = -D\co{\phi},
\end{align}
in the Campbell-Baker-Hausdorff formula, where these equations are
meant to act on forms, we also find that
\begin{equation}
\label{cbh}
e^{D\co{\phi}} e^{-\co{\theta}} = e^{-\co{\theta}} e^{\frac{1}{\theta}}
  e^{\co{\theta D\phi}} e^{-\frac{1}{\theta}},
\end{equation}
where $(\theta D\phi)^{\mu i} = \theta^{\mu \nu} D_{\nu} \phi^i$.
Using these formulas, we obtain
\begin{alignat}{2}
&&& \int \left(e^{-\co{\theta}} e^{f} \Pb\right)
  e^{-i \co{\com{\phi}{\phi}}}  C e^B \tag{\ref{rrf}} \\
&\text{\small{[via~\eqref{altPb},~\eqref{moveco}]}}
&& = \int  e^{f} e^{D\co{\phi}} e^{-\co{\theta}} 
  e^{-i \co{\com{\phi}{\phi}}}  C e^B 
\nonumber \\
\label{alfp}
&\text{\small[via~\eqref{cbh}]}
&&=
\int   e^{f}  e^{-\co{\theta}} e^{\frac{1}{\theta}}
  e^{\co{\theta D\phi}} e^{-\frac{1}{\theta}}
  e^{-i \co{\com{\phi}{\phi}}}  C e^B 
\\
&\text{\small[via~\eqref{intc}]} 
&&= 
  (-1)^{\frac{p(p+1)}{2}}
\int \Pf(f) \, \star \left[ e^{\co{f^{-1}-\theta}}
  e^{\frac{1}{\theta}} e^{\co{\theta D\phi}} 
  e^{-i \co{\com{\phi}{\phi}}}  C e^{B-\frac{1}{\theta}} \right] 
\nonumber \\
\label{mukhi2}
&\text{\small[via~\eqref{intc}]}
&&= 
\int  \frac{\Pf(\theta-\theta f \theta)}{\Pf(\theta)}
   e^{\frac{1}{\theta-\theta f \theta}}
   e^{\co{\theta D\phi}} 
  e^{-i \co{\com{\phi}{\phi}}} C e^{B-\frac{1}{\theta}}
\\ 
&\text{\small[via~\eqref{intc}]} 
&&= 
\frac{1}{\Pf(\theta)} \int 
 \star\left(  e^{\co{\theta-\theta f \theta}}
   e^{\co{\theta D \phi}} 
  e^{-i \co{\com{\phi}{\phi}}}  C e^{B-\frac{1}{\theta}} \right)
\nonumber \\
&\text{\small[via~\eqref{defmf}]}
&&  = 
\frac{1}{\Pf(\theta)}
 \int  \star \left[ e^{-i \co{\com{X}{X}}} C 
e^{B-\frac{1}{\theta}} \right],
\end{alignat}
where we have used the identities \hbox{$\frac{1}{\Pf(-f^{-1})} = \Pf(f)$}
and \hbox{$\Pf(\theta-f^{-1}) \Pf(\theta) \Pf(f) = \Pf(\theta-\theta
f\theta)$}.%
\iftoomuchdetail%
\footnote{The last identity can be proven by noting that (in even
dimension) (a) the square of it is an obvious determinant identity, and
(b) it
is true for $2\times
2$---and therefore any simultaneously skew diagonalizable---matrices $f$ and
$\theta$.  The first means that it is true up to a sign.  Since the
identity is
a polynomial equation in the matrix
elements, the second means
that the sign must also be correct by analyticity.}%
\fi\ 
The above manipulations use the commutative product structure 
between various open string fields, which applies to~\eqref{rrf}
under the symmetrized trace $\STr$, and generally follows from full
symmetry inside $L_{\ast}$.

Now we give another form of~\eqref{rrf} and its finite momentum 
version~\eqref{rrfw}, which is also very useful. 
Using~\eqref{moveco} and~\eqref{alfp} we find that
\begin{equation} \label{trr}
\begin{split}
\swz & = 
         \mu_p \, \STr \int \left( e^{- \co{\theta}} e^{f} \right) \left(
e^{\frac{1}{\theta}} \, e^{\co{\theta D \phi}} \, 
e^{-i \co{\com{\phi}{\phi}}} \;
e^{B -\frac{1}{\theta}} \; C \right) \\
& = \mu_p  \STr \int \sqrt{\det (1 - \theta f)} \; 
e^{f \frac{1}{1- \theta f}}
\left( e^{\frac{1}{\theta}} \, e^{\co{\theta D \phi}} \, 
e^{-i \co{\com{\phi}{\phi}}} \; e^{B -\frac{1}{\theta}} \; C \right)
\end{split}
\end{equation}
Note that now $e^{-\co{\theta}}$  acts only on $e^{f}$.
In the second line above we have used an identity 
\begin{equation} \label{ith} 
e^{-\co{\theta}} e^{f} = 
\sqrt{\det (1 - \theta f)} \; e^{f \frac{1}{1- \theta f}}.
\end{equation}
The identity is derived using~\eqref{itow}:
\begin{multline}
e^{-\co{\theta}} e^{f}
=  e^{-\co{\theta}} \star\left[(\star e^{f})(1)\right]
= (-1)^{\frac{p(p+1)}{2}}
  \Pf(f) e^{-\co{\theta}} e^{\co{f^{-1}}} (\star 1) \\
= (-1)^{\frac{p(p+1)}{2}}
  \Pf(f) \Pf(f^{-1}-\theta) \star \left[(\star
e^{\frac{1}{f^{-1}-\theta}}) (1)\right]
= \sqrt{\det(1-\theta f)} e^{f \frac{1}{1-\theta f}}.
\end{multline}
In the last step we take $f$ to be sufficiently small relative to
$\theta^{-1}$ so that the sign is unambiguous.

To summarize, with the scalar fields set to zero we have three
equivalent expressions that yield the charge coupling to the \RR\ fields, 
namely,
\begin{equation}
\begin{split}
\swz & = 
   \mu_p \, \STr \int \left( e^{- \co{\theta}} e^{f} \right)
\;  C e^B \\
& = \mu_p \, \STr \int \sqrt{\det (1 - \theta f)} \; 
e^{f \frac{1}{1-\theta f}} C e^B 
\\
& = 
\mu_p \, \STr \int  \star \left[ e^{-i \co{\com{X}{X}}} \, 
e^{-\frac{1}{\theta}} \, C e^{B} \right]
\end{split}
\end{equation}
Again we emphasize that the \RR\ charge of a gauge configuration 
should be measured using $C e^B$.

Finally note that the open Wilson line~\eqref{cwilson} can be written in
Matrix Theory variables~\eqref{ncb} as~\cite{wadia}
\begin{equation} 
W(x, \cC_q)\ast e^{iqx} = \exp \left(i q \cdot X \right) = \exp 
\left(i q_{\mu}  X^{\mu} + i q_i X^i \right)
\end{equation}
Using this, the complete finite momentum action can now be written as 
\begin{equation} \label{mmq}
\swz =
\mu_p  \frac{1}{\Pf(\theta)} L_\ast \int d^{10} q \star 
\left(e^{-i \co{\com{X}{X}}} \, e^{-\frac{1}{\theta}}\,  C(q) e^{B}\,  
e^{i q \cdot X} \right).
\end{equation}

\section{An Analysis of Higher Powers} \label{sec:wepb}

In the previous section, we have verified the first few terms
of~\eqref{rrf} to the amplitudes derived in section~\ref{sec:amp}.
Here we will verify some of the higher order terms.  Specifically, for
the terms that we have found in~\eqref{amp0}, we
can verify the presence of the Wilson line and the $L_{\ast}$ prescription 
to all orders in the gauge field.  We can also verify the presence of 
Elliott's formula in the
action to all orders in the gauge field, albeit without the
interaction term in the field strength.  That is, we can verify
$\left(e^{-\co{\sfrac{\theta}{2\pi\apr}}} e^{2\pi\apr da}\right) C
e^B$.
This verifies the
zero-momentum action~\eqref{rrf} to all orders in the gauge field, in
the absence of transverse scalars, assuming the completion of $da$ to
$f$.  This is a reasonable assumption by gauge invariance.
We can also incorporate, to all
orders, the pullback in equation~\eqref{rrf}, although again our
computation is not sensitive to the presence of covariant, rather than
partial, derivatives in the definition of the pullback.

Note that none of the aforementioned verifications involve
interactions of the gauge field amongst themselves.  Those are more
complicated to verify.  In particular, the Myers term is an interaction term,
so we are, unfortunately, not going to verify it to higher order (for the 
commutative theory, some higher order Myers terms were checked in~\cite{gm2}).

\subsection{The Wilson Line} \label{sec:w}

In this subsection we verify that
the terms in the quadratic action,~\eqref{amp0},
are attached to an open Wilson line with the 
$L_{\ast}$  prescription.
This will demonstrate equation~\eqref{rrfw} to quadratic order.
Our strategy is to isolate, from the amplitude
with $n$ open strings, a particular subset of terms which 
will be identified with the Wilson line completion of~\eqref{amp0}. 
The discussion of this subsection will closely 
parallel section 4 of~\cite{lm3}.%
\footnote{See also~\cite{ooguri} for a nice discussion of 
the Wilson line structure from the amplitudes in the $\apr \rightarrow 0$ 
limit. Our discussions here and in~\cite{lm3} are slightly more general.}

For an amplitude with $n$ open strings~\eqref{namp}, we can split off one or
two open strings and obtain correlation functions like those
that gave rise to~\eqref{founda1} or~\eqref{twoamp}.  For the
remaining factor, we consider the subset of terms that involve
$\dot{X}$; specifically, those that give rise to factors of the
form~\eqref{bn}.  In particular, we focus on the middle two terms,
\begin{equation} \label{xdotma}
\frac{1}{2\pi\apr}\frac{1}{1+y_a^2} \cM_a%
; \qquad \cM_a = i(q\times a)+ i (2\pi\apr) a\cdot q_\perp.
\end{equation}
Na\"{\i}vely, this, 
along with the overall exponential phase factor 
\begin{equation} \label{kernel}
 \exp\left[\sum_{a<b}\frac{i}{2} 
\bigl(k_a \times k_b\bigr) \bigl(2 \tau_{ab} -
\epsilon(\tau_{ab})\bigr)
\right],
\end{equation}
from~\eqref{corr1} in the $\apr k_a \cdot k_b\rightarrow 0$ limit, is
precisely what is required to reproduce the Wilson line.
Specifically, the factors of $\frac{1}{1+y_a^2}= \pi \frac{d \tau_a}{dy_a}$ 
give the measure for
the change of variables $y_a = -\cot(\pi \tau_a)$; then the integration
of~\eqref{kernel}
gives the
$\ast_n$ $n$-ary operation that we expect from the expansion of the
Wilson line~\cite{liu}.
Thus,~\eqref{xdotma} essentially shows that we obtain the Wilson line;
the numerical factor for the exponentiation follows from the
combinatorics in converting an amplitude to an action.

Strictly speaking, however, in order for this to work properly,
we have to make sure that $y_1$ and $y_2$ also appear properly.
In particular, because poles can appear in the amplitudes---or
equivalently, because the amplitudes are generally defined by analytic
continuation of the $y_a$ integrals---it is generally troublesome to
take the $\apr\rightarrow0$ limit before integrating.  However,
as in~\cite{lm3},
one can check that the integrand is regular if we take all $\apr k_a\cdot
k_b\rightarrow 0 $ except $\apr k_1\cdot k_2$.  This leaves, in
addition to the contribution from~\eqref{kernel}, an overall
trigonometric factor involving $\tau_2$ (we fix $\tau_1=\thalf$),
which is generically singular as $\apr k_1\cdot k_2\rightarrow 0$, but
can be made nonsingular, except for a possible explicit factor of
$\frac{1}{\apr k_1\cdot k_2}$, via an integration by parts.  Then we can also
take $\apr k_1\cdot k_2 \rightarrow 0$, and throwing out the explicit poles,
recover the $\ast_n$ kernel~\eqref{kernel}.

For example, for the generalization of equation~\eqref{twoamp} to $n$ open
strings,
the terms of interest, after taking the $\apr k_a\cdot k_b\rightarrow
0$ limit,  are of the form
\begin{equation} \label{nopen}
\int_0^1 d\tau_2 \cdots \int_0^1 d\tau_n
\prod_{a<b}
\exp\left[\frac{i}{2} 
\bigl(k_a \times k_b\bigr) \bigl(2 \tau_{ab} -
\epsilon(\tau_{ab})\bigr)
\right] \, \abs{\cos(\pi \tau_2)}^{2\apr k_1\cdot k_2} \, \sT \,   
\cM_3\dots \cM_n
\end{equation}
where $\sT$ denotes the integrand of~\eqref{twoamp} with $A_2$ removed, i.e.
the sum of~\eqref{i00}--\eqref{I44} with the factor $A_{2} (y_2)$ removed. 
There are three types of terms in $\sT$ classified  according to 
their dependence on $y_2$ or $\tau_2$, which we shall now analyze one by one:

\begin{description}

\item[Terms depending only on $\frac{1}{1 + y_2^2}$:] 

These terms give rise to the
contact terms in our discussion in section~\ref{sec:amp}. Since there is no
singularity in $y$, we can take the limit $\apr k_1 \cdot k_2 \rightarrow 0$ 
in the  integrand and the integrations~\eqref{nopen} precisely give $\ast_n$  
operations, which corresponds to the affixation of an open Wilson line with
$L_{\ast}$ ordering. 

\item[Terms singular at $y_2 = 0$ and without an explicit
$\apr k_1 \cdot k_2$:] 

These terms gave rise 
to pole terms in section~\ref{sec:amp} and were discarded. It is easy to argue 
that they also contain poles in~\eqref{nopen}. The idea is that we can
perform an 
integration by parts in the integrand so that the singularity in $y_2$ is 
replaced by a pole in $\apr k_1 \cdot k_2$. For example to take care
of a simple pole in $y_2$---or a factor of $\tan \pi \tau_2$---%
we use the identity
\begin{equation} \label{trick}
\abs{\cos(\pi \tau_2)}^{2\apr k_1\cdot k_2} \tan(\pi \tau_2)
= -\frac{1}{2\pi\apr k_1\cdot k_2} \frac{\p}{\p \tau_2}
 \abs{\cos(\pi\tau_2)}^{2\apr k_1\cdot k_2}.
\end{equation}
Since the remainder of the integrand is nonsingular, we can integrate by
parts, and find that the resulting integrand is $\frac{1}{2\pi\apr
k_1\cdot k_2}$ times a nonsingular integrand. 

\item[Terms singular at $y_2 = 0$ and with an explicit
$\apr k_1 \cdot k_2$:]

In section~\ref{sec:amp}, these terms gave rise 
to contact terms, thereby yielding the
$[a_{M}, a_{N}]_{\ast}$ completion
to the noncommutative field strength, (covariant) pullback, and Myers 
type terms. Using the same trick~\eqref{trick}, the singular part in the 
integrand can again be written as a pole in  $\apr k_1 \cdot k_2$
times a nonsingular integrand, for which we can take  the 
limit $\apr k_1 \cdot k_2 \rightarrow 0$ before doing the integration.
Now the pole precisely cancels with the multiplicative factor 
$\apr k_1 \cdot k_2$ and gives rise to a contact term.
A careful integration by parts yields
precisely
$[a_{M}, a_{N}]_{\ast}$ attached to a Wilson line.
Note that if we had na\"{\i}vely taken $\apr k_a \cdot k_b\rightarrow 0$
inside the integrals,
we would have missed the contributions of these terms.

As the procedure just outlined is rather subtle, we will
demonstrate this explicitly.
The requisite integration, after employing the trick~\eqref{trick}, is
\begin{equation} \label{intm*}
I = \int_0^1 d\tau_n \cdots \int_0^1 d\tau_2 
\left(\frac{\p}{\p\tau_2} \abs{\cos{\pi \tau_2}}^{2\apr k_1\cdot k_2}\right)
\exp\left[\frac{i}{2} \sum_{a<b} (k_a \times k_b)\bigl(2 \tau_{ab} - 
   \epsilon(\tau_{ab})\bigr)\right];
\end{equation}
recall that $\tau_1=\frac{1}{2}$.
We now integrate by parts.  Note that there are many surface terms,
since pointsplitting regularization on the worldsheet implies that we
exclude the points $\tau_2 = \tau_a$ (for $a\neq 2$) from the integration.
So we have
\begin{multline}
\!\!
I = \int_0^1 d\tau_n \cdots \int_0^1 d\tau_3
\sum_{\tau_2 \in {\mathcal S}} s_{\tau_2}
\abs{\cos{\pi \tau_2}}^{2\apr k_1\cdot k_2}
\exp\left[\frac{i}{2} \sum_{a<b} (k_a \times k_b)\bigl(2 \tau_{ab} - 
   \epsilon(\tau_{ab})\bigr)\right]
\\
-  \int_0^1 d\tau_n \cdots \int_0^1 d\tau_2
\abs{\cos{\pi \tau_2}}^{2\apr k_1\cdot k_2}
\frac{\p}{\p\tau_2} 
\exp\left[\frac{i}{2} \sum_{a<b} (k_a \times k_b)\bigl(2 \tau_{ab} - 
   \epsilon(\tau_{ab})\bigr)\right].
\end{multline}
with the first term the surface contributions. Here ${\mathcal S}$ is the set
$\{0,\frac{1}{2}^{\mp},\tau_3^{\mp},\cdots,\tau_n^{\mp},1\}$ (recall that 
$\tau_1 =\ha$), where, of course, the superscript denotes whether we approach 
from above or below, and the signs $s_{\tau_2}$ are
\hbox{$s_{\tau_2=\tau_a^{\mp}}$ = $\pm 1$}, \hbox{$s_{\tau_2=0}=-1$}
and \hbox{$s_{\tau_2=1}=1$}.
Since the $\tau_2$ integration in the second term above 
is now nonsingular, we can take $\apr
k_1\cdot k_2 \rightarrow 0$ in the integrand thereby obtaining a total
derivative.  So we now have,
\begin{multline} \label{you}
I = \int_0^1 d\tau_n \cdots \int_0^1 d\tau_3
\sum_{\tau_2 \in {\mathcal S}} s_{\tau_2}
\left\{\abs{\cos{\pi \tau_2}}^{2\apr k_1\cdot k_2} - 1\right\}
\\ \times
\exp\left[\frac{i}{2} \sum_{a<b} (k_a \times k_b)\bigl(2 \tau_{ab} - 
   \epsilon(\tau_{ab})\bigr)\right].
\end{multline}
Of course, away from the zeros of the cosine---that is, away from
$\tau_2=\frac{1}{2}^{\mp}$---the quantity in curly brackets vanishes
as $2\apr k_1\cdot k_2\rightarrow 0$.  Thus we are left with
the contribution from
$\tau_2 = (\ha)^{\mp}$ in~\eqref{you},
\begin{multline} \label{fintm*}
I = 2i \sin \frac{k_1\times k_2}{2}
\int_0^1 d\tau_n \cdots \int_0^1 d\tau_3
\\
\exp\left[ \frac{i}{2} \sum_a [(k_1+k_2)\times k_a] \bigl(2 \tau_{1a}
- \epsilon(\tau_{1a})\bigr)
+ \frac{i}{2} \sum_{2<a<b} (k_a \times k_b)\bigl(2 \tau_{ab} - 
   \epsilon(\tau_{ab})\bigr)\right].
\end{multline}
We recognize that the integration in~\eqref{fintm*} gives
$\ast_{n-1}$;
thus we obtain
\hbox{$\ast_{n-1}\Bigl[\com{a_{1M}(k_1)}{a_{2N}(k_2)}_\ast,
\cM_3(k_3),\cdots,\cM_n(k_n)\Bigr]$}.
This completes the
demonstration; the reader can check that the numerical factors work
properly.  
Again note that taking $\apr k_1\cdot k_2\rightarrow
0$ prematurely would have resulted in the omission of this important term.
\end{description}

\subsection{Elliott's Formula and The Pullback} \label{sec:elliot}

In this subsection we check the couplings~\eqref{rrfw} to all 
orders up to the nonlinear terms in the field strengths, i.e.\ up to 
terms of the form $\com{a_{M}}{a_{N}}_\ast$.

The simplest correlation functions are those for which we ignore both the
$\dot{X}$ part of the vertex operator~\eqref{vopen}, and the Wick
contractions~\eqref{wick}.  Then the
correlation function simply gives (suppressing the $\tau_a$
integrations and setting $\mu_p \kappa_{10} = 1$)
\begin{equation} \label{justpsi}
\frac{\lambda}{2i} 
(2\pi i\apr)^n A_n k_{1\mu_1} e_{1M_1} \dots k_{n\mu_n} e_{nM_n}
\Lambda^{\mu_1 M_1
\dots \mu_n M_n}.
\end{equation}
As in the previous subsection, in the 
$\apr k_a\cdot k_b\rightarrow0$ limit, the
factor of $A_n$ gives rise precisely to the $\ast_n$ $n$-ary operation between
the open string modes in the expression;
there is no subtlety here because the only $y_a$-dependence is that in
equation~\eqref{kernel}.

If we take only longitudinal polarizations in
equation~\eqref{justpsi},
(and take the low energy limit) then
equation~\eqref{expgiveL} gives us
\iftoomuchdetail
(there is an extra factor of $2^{-n}$ from $k e = \half da$, but note that
$f^n = \frac{1}{2^n} f_{\mu_1\mu_2} \cdots f_{\mu_{2n-1}\mu_{2n}}
dx^{\mu_1} \cdots dx^{\mu_{2n}}$)
\fi
\begin{equation} \label{preelliot}
(2\pi\apr)^n \star \left[ \left(e^{-\co{\sfrac{\theta}{2\pi\apr}}} 
\ast_n[da_1, \cdots,
 da_n]\right)
 C e^B \right]_{\text{0-form}}
\end{equation}
where the wedge product of the field strengths is combined with the
$\ast_n$ $n$-ary operation.  Summing over $n$, with a $1/n!$
indistinguishability factor, gives Elliott's formula,
\begin{equation}
\left( e^{-\co{\sfrac{\theta}{2\pi\apr}}} e^{2\pi\apr f} \right) C e^B.
\end{equation}
as given in~\eqref{rrfw}.  Here we have made the replacement
$da\rightarrow f$; the extra $a\ast a$ terms
should come from a Wick contraction that we ignored, as it did for
equation~\eqref{I42}---see, in particular, the comments just preceding
equation~\eqref{ff}.

If we take the open string modes in~\eqref{justpsi} to be purely
transverse, then~\eqref{expgiveL} gives
\begin{equation} \label{prepb}
(2\pi\apr)^n \star \left[ \left(e^{-\co{\sfrac{\theta}{2\pi\apr}}} 
\ast_n[d\phi_1^{i_1}, \dots,
 d\phi_n^{i_n}]\right)
 C_{i_1\dots i_n} e^B \right]_{\text{0-form}}
\end{equation}
where the longitudinal indices are suppressed and treated as forms.
\iftoomuchdetail
The $n!$ in equation~\eqref{expgiveL} is canceled by the $n!$ in the
definition of $\co{T}$.
\fi
On replacing the partial derivatives with covariant derivatives, this
precisely reproduces the $n$th term of the pullback of $Ce^B$, as given in
equation~\eqref{rrfw}.

More generally, we can take $q$ of the open string modes to be
longitudinal and $r$ to be transverse.  Then we find
\begin{equation} \label{preellpb}
(2\pi\apr)^{q+r} \star \left[ \left(e^{-\co{\sfrac{\theta}{2\pi\apr}}}
\ast_{q+r}[da_{1}, \dots, da_q,
 d\phi_1^{i_1}, \dots,
 d\phi_r^{i_r}]\right)
 C_{i_1\dots i_r} e^B \right]_{\text{0-form}}.
\end{equation}
Summing over $q$ and $r$, and including the indistinguishability
factors, exponentiates these precisely to give an action
\begin{equation} \label{ellandpb}
\int \left[e^{-\co{\sfrac{\theta}{2\pi\apr}}} e^{da} \Pb_\p\right] C e^B,
\end{equation}
where the subscript on the pullback is to emphasize that here we do not
have the covariant derivative in the definition of the pullback.  
Also, we have suppressed the $n$-ary operation.
This differs
from~\eqref{rrfw} only by pure open string interaction terms $a\ast a$,
$a \ast \phi$ and $\phi \ast \phi$, and the explicit inclusion of the
Wilson line.  Of course, we checked the Wilson line for the first
couple of open-string powers in the previous subsection.

\section{Conclusion} \label{sec:conc}

We believe we have presented convincing evidence that equation~\eqref{rrfw}
gives the \WZ\ terms for noncommutative D-branes.  In
particular, it precisely matches amplitude calculations to quadratic
order in the gauge field, as shown in section~\ref{sec:wz}.
Furthermore, in section~\ref{sec:wepb}, we have found the Wilson line
to all orders in the gauge field.  We have confirmed the presence,
in the action, of both Elliott's formula and the pullback, up to
interaction terms in the field strength and covariant derivatives.
We have also shown that our result for the \WZ\ term that we have
given here, agrees with that derived from the
Matrix model.

While we have not discussed it explicitly in the paper, it is a simple 
matter to check that the on-shell amplitudes we computed in 
section~\ref{sec:amp} are invariant under a gauge transformation of 
the \RR\ potential. However to check that the off-shell extension we 
proposed in section~\ref{sec:wz} is gauge invariant---or equivalently,
that the expression coupled to $C$ in equation~\eqref{rrfw} is
closed---is 
rather 
complicated.
In~\cite{oo}, it was argued that the action, written in the Matrix
model form~\eqref{mmq}, is gauge 
invariant. Also, some simple limits 
of equation~\eqref{rrfw} are obviously gauge invariant.
For example, in the
commutative limit it is known to be gauge invariant.
Also, Elliott's
formula is closed, thereby implying gauge invariance 
at zero momentum, in the absence of
transverse scalar fields.  

We should note that although the \WZ\ term we have given here
involves the $B$-field, in principle this is merely the background
value of the $B$-field; we have no guarantee that
equation~\eqref{rrfw} properly
incorporates fluctuations of the $B$-field.  This is because we have
only computed disk amplitudes with the insertion of one closed string
mode.
It is interesting to note
that, because the commutative limit of equation~\eqref{rrfw} precisely
reproduces the commutative \WZ\ term, in this limit fluctuations of
$B$ are incorporated in the action.  However, there is a general
argument~\cite{oow} that the Wilson line prescription for couplings of
two or more closed string modes to an open string operator is more
complicated than that in~\eqref{rrfw}. It would be interesting to 
analyze the problem of $B$-field fluctuations more thoroughly. 

\acknowledgments

We are grateful for useful conversations with C.~Hofman, G.~Moore,
L.~Motl and H.~Ooguri.
This work was supported in part by DOE grant
\hbox{\#DE-FG02-96ER40559}.  J.M. was also supported by an NSERC PDF
Fellowship and Grant
No. PHY99-07949 of the National Science Foundation.

\appendix 

\section{On the RR Vertex Operator} \label{app:C*C}

We wish to write down a ($-1/2,-3/2$)-picture vertex operator for the
\RR\ fields.
In~\cite{divech}, it was shown that, under picture-changing,
the known ($-1/2,-1/2$)-picture
operator is reproduced by the vertex operator
\begin{equation}
V^{-1/2, -3/2}_{\text{RR}} = \frac{2 g_c}{\apr} e^{- \phi/2 - 3\tilde{\phi}/2}
\Theta \ch \frac{\one-\Gamma^{11}}{2} \fs{C^{(n)}} \tilde{\Theta} e^{i q \cdot
X},
\end{equation}
but that this $V^{-1/2,-3/2}_{\text{RR}}$ is BRST closed only when
$dC^{(n)}=d\hat{\star} C^{(n)}=0$; i.e.~only for
vanishing field strengths.  Here and throughout this paper, $\hat{\star}$
is the 10-dimensional Hodge dual with respect to the closed string
metric.
If we add in another \RR\
potential, and write
\begin{equation}
V^{-1/2, -3/2}_{\text{RR}} = \frac{2 g_c}{\apr} e^{- \phi/2 - 3\tilde{\phi}/2}
\Theta \ch \frac{\one-\Gamma^{11}}{2} 
\left[\fs{C^{(n)}}+\fs{C^{(n+2)}}\right]
\tilde{\Theta} e^{i q \cdot
X},
\end{equation}
then BRST invariance requires~\cite{divech}%
\iftoomuchdetail%
\footnote{In \cite{divech}, the middle equation differs by a relative
sign, so we would like to justify the sign here.  The point is that
the equation is really~\cite{divech} $\fs{k}
(\fs{C^{(n)}}+\fs{C^{(n+2)}})=0$.  With the conventions for the
$\Gamma$-matrices given here, $\fs{k} \fs{C^{(n)}}$ is 
(roughly---we omit numerical
factors of positive sign) $dC^{(n)} + \p\cdot C^{(n)}$ where $\p\cdot
C^{(1)}$ = $\p^\mu C_\mu$, etc.  On a $2d$-dimensional, Lorentzian
space, $\p\cdot = -\hat{\star} d \hat{\star}$.  Thus there is a relative
minus sign.}%
\fi
\begin{align} \label{brst2}
d\hat{\star} C^{(n)} &= 0, &
d C^{(n)} - \hat{\star} d \hat{\star} C^{(n+2)} &= 0, &
d C^{(n+2)} &= 0.
\end{align}
Via a gauge transformation, $\delta C^{(n+2)} = d\Lambda^{(n+1)}$, we
can always satisfy the middle equation (since one can always solve the
``Poisson'' equation); thus we now have an \RR\ vertex
operator for arbitrary $F^{(n+1)} = dC^{(n)}$ field strength.  So,
by adding in all the \RR\ potentials as in~\eqref{vclose}, we have a
BRST invariant vertex operator for arbitrary field strengths.  The
analogues of the first and last equations of~\eqref{brst2} are
automatically satisfied since there are no eleven-forms in a
ten-dimensional theory.%
\footnote{For IIA, we note that $F^{(10)}$ always vanishes in
perturbation theory.  
}

We note that the operator $\half (\one-\Gamma^{11})$ imposes a chirality
condition on the spinors which then projects out the anti-self dual
part of $C=\sum C^{(n)}$, keeping only 
$\Gamma^{11}\fs{C} = -\fs{C}$.
This is a self duality condition since on an $n$-form $\omega^{(n)}$,
\begin{equation} \label{g11->*}
\Gamma^{11} \fs{\omega^{(n)}} = (-1)^{\frac{n(n+1)}{2}}
\fs{\hat{\star} \omega^{(n)}}.
\end{equation}
It is convenient to include this 
sign as part of the definition of self vs.\ anti-self duality.  Note
also that the closed string metric arises because of the closed string
$\Gamma$-matrices.

One might wonder about the relationship---or even the
compatibility---of self duality of $C$
and the statement from
supergravity---or from the $(-1/2,-1/2)$-picture---that the on-shell field
strengths are self dual:
$\Gamma^{11} \fs{F} = -\fs{F}$.%
\footnote{Note that with this sign, equation~\eqref{g11->*} indeed
gives $F^{(5)} = \hat{\star} F^{(5)}$.
\iftoomuchdetail%
The reader who would prefer
e.g.\ $F^{(3)}=\hat{\star} F^{(7)}$ to the $F^{(3)}=-\hat{\star}
F^{(7)}$ that we have here, need merely redefine $F^{(7)}$ (and
$C^{(6)}$) by a sign.  
\fi%
Note that since $(\Gamma^{11})^2 = \one$,
the (anti-)self duality of even-dimensional forms is well-defined
here.\label{ft:dual}} \ 
It turns out that self duality of the field strength is
equivalent (up to a gauge
transformation) to self duality of the potential
in the gauge implied by the generalization
of~\eqref{brst2},
\begin{equation} \label{fbrst}
d C^{(n)} - \hat{\star} d \hat{\star} C^{(n+2)} = 0.
\end{equation}

In particular, we want to show that self duality of $C$ implies self
duality of $F$.  So suppose that $C$ is self dual.  Then,
\begin{equation}
F^{(n)} = dC^{(n-1)} = -(-1)^{\frac{n(n+1)}{2}} d\hat{\star} C^{(11-n)}
= (-1)^{\frac{n(n-1)}{2}} \hat{\star} d C^{(9-n)} 
= (-1)^{\frac{n(n-1)}{2}} \hat{\star} F^{(10-n)},
\end{equation}
where in the second step we used self duality of $C$ and in the third
step, we used~\eqref{fbrst}.
Thus, we see that $F$ is self dual: 
$F^{(10-n)} = -(-1)^{\frac{n(n+1)}{2}} \hat{\star} F^{(n)} 
   \Rightarrow \Gamma^{11} \fs{F} = -\fs{F}$ by
equation~\eqref{g11->*}.

Conversely, suppose the field strength is self dual.
Then,
\begin{equation} \label{F*FtoC*C}
dC^{(n)} = F^{(n+1)} 
         = (-1)^{\frac{n(n+1)}{2}} \hat{\star} F^{(9-n)}
         = (-1)^{\frac{n(n+1)}{2}} \hat{\star} dC^{(8-n)}
         = (-1)^{\frac{n(n-1)}{2}} d\hat{\star} C^{(10-n)},
\end{equation}
and so
$C^{(n)} = (-1)^{\frac{n(n-1)}{2}} d\hat{\star} C^{(10-n)}$, up to an
exact form (i.e.~a gauge
transformation%
\iftoomuchdetail%
, as closed forms are exact on ${\mathbb{R}}^{10}$%
\fi%
).
This is equivalent to $\Gamma^{11} \fs{C} = -\fs{C}$ and completes the proof.

Note that while the vertex operator implies a special choice of
gauge for $C^{(n)}$, the final on-shell amplitudes are gauge invariant and 
do not depend on the choice of the gauge.    

\section{A Derivation of the Fermionic Boundary Condition} \label{app:m}

We will show that for a D$p$-brane with $B$-field background, the 
worldsheet boundary conditions for the spin operators are
$\tilde{\Theta}(\bar{z})=M\Theta(z)$ with $M$ given by
\begin{equation} \tag{\ref{mforrl}}
M = \frac{\sqrt{-\det g}}{\sqrt{-\det{(g+B)}}} 
\AE\left(\fs{B}\right) \Gamma^0 \dots
\Gamma^{p} \begin{cases} \one, \quad& \text{type IIA} \\
                                   \Gamma^{11}, & \text{type IIB} \end{cases}.
\end{equation}
$\AE(\fs{B})$, as defined in~\cite{callan}, is the
exponential of $\fs{B}$, but with the $\Gamma$ matrices totally
antisymmetrized.  (This is equivalent to $\fs{e^B}$, where wedge
products are understood in the definition of the exponential, and
where, as in~\eqref{fs}, the Feynman slash of a sum is the sum of
Feynman slashes.) Our derivation of~\eqref{mforrl} here generalizes those
in~\cite{malm,myersm} to non-zero $B$ (see also~\cite{witd0}).

The consistency of the OPEs
\begin{align} \label{opett}
\Theta_A(z_1) \Theta_B(z_2) &\sim \frac{\ch^{-1}_{AB}}{z_{12}^{5/4}}, 
& \tilde \Theta_A(\bar z_1) \tilde \Theta_B(\bar z_2) & 
\sim \frac{\ch^{-1}_{AB}}{\bar z_{12}^{5/4}} \\
\label{opept}
\psi^{M}(z_1) \Theta_A(z_2) &\sim 
\frac{\Gamma^M_{AB} \Theta_B(z_2)}{2 z_{12}^{1/2}},
& \tilde \psi^{M}(\bar z_1) \tilde \Theta_A(\bar z_2) &\sim 
\frac{\Gamma^M_{AB} \tilde \Theta_B(\bar z_2)}{2 \bar z_{12}^{1/2}}
\end{align}
with the boundary conditions at $z = \bar z$
\begin{equation}
\tilde{\Theta}(\bar{z})=M\Theta(z), \quad
\tilde{\psi}^\mu(\bar{z}) = 
   \left(\frac{g+B}{g-B}\right)^{\mu}{_\nu} \psi^\nu(\bar{z}), \quad
\tilde{\psi}^i(\bar{z}) = -\psi^i(\bar{z}),
\end{equation}
requires that $M$ satisfies 
\begin{gather} \label{mcondc}
\ch^{-1} = M \ch^{-1} M^\transpose, \\
\label{mcondg}
\begin{align}
\Gamma^{\mu} M &= \left(\frac{g+B}{g-B}\right)^{\mu}{_\nu} M \Gamma^\nu, &
\Gamma^i M &= -M \Gamma^i,
\end{align}
\end{gather}
where $\ch_{AB}$ is the charge conjugation matrix, for which
\begin{equation} \label{defcconj}
\ch \Gamma^\mu \ch^{-1} = -\left(\Gamma^\mu\right)^\transpose.
\end{equation}

Equation~\eqref{mcondg} states that $M$ is the action, in a spinor basis,
of the transformation which acts as a Lorentz rotation 
$ \left(\frac{g+B}{g-B}\right)^{\mu}{_\nu}$ in the longitudinal directions
and an inversion of all coordinates in the transverse directions. 
Since $B$ only affects the longitudinal directions, we decompose 
$M = M_0 S$, where $M_0$ is a solution of~\eqref{mcondc} and~\eqref{mcondg}
with $B=0$ and $S$ acts on spinors as a Lorentz transformation 
$ \left(\frac{g+B}{g-B}\right)^{\mu}{_\nu}$ in the longitudinal directions.
The expression for $M_0$ is well-known, and is given by
\begin{equation}
M_0 = \Gamma^0 \cdots \Gamma^{p} \begin{cases} \one, & \text{IIA} \\
 \Gamma^{11}, & \text{IIB} \end{cases},
\end{equation}
while $S$ has the form
\begin{equation}
S_{AB} = \exp \left(\frac{i}{2} \omega_{\mu \nu} \Sigma^{\mu \nu} \right)
\end{equation}
where $\Sigma^{\mu \nu} = - \frac{i}{4} \com{\Gamma^{\mu}}{\Gamma^{\nu}}$
are  Lorentz generators in the spinor basis. 
Since $S$ always contains an even number of longitudinal 
$\Gamma$-matrices, it satisfies 
\begin{equation} \label{conS}
\ch^{-1} = S \ch^{-1} S^\transpose, \qquad
\Gamma^i S = S \Gamma^i.
\end{equation}
and commutes with $M_0$. Thus we have shown that $M=M_0 S$  
satisfies~\eqref{mcondc} and~\eqref{mcondg}.

To find the explicit form of $S$ in terms of $B$, it is convenient to 
choose a basis so that $g_{\mu \nu} = \eta_{\mu \nu}$ and 
$B$ is skew diagonal. Without loss of generality
let us look at a Euclidean two-dimensional block $a,b =2,3$ with 
$B = \bigl(\begin{smallmatrix}
0 & \lam \\ -\lam & 0
\end{smallmatrix} \bigr)$.  Then 
\begin{equation}
 \left(\frac{1}{g-B}\right) (g+B)
= \begin{pmatrix}
\cos \theta &  \sin \theta \\ 
- \sin \theta & \cos \theta 
\end{pmatrix}
\end{equation}
with $\lam= \tan \frac{\theta}{2}$. Now we can immediately write down 
the transformation in the spinor basis,
\begin{equation}
\begin{split}
S & = \exp \left[\ha \theta \, \Gamma^{2} \Gamma^{3} \right] \\
& =\cos \frac{\theta}{2} + \sin \frac{\theta}{2}\, \Gamma^{2} \Gamma^{3} \\
& = \frac{1}{\sqrt{\det (\eta+B)}} 
(1 + \ha B_{a b} \Gamma^{a}\Gamma^{b})
\end{split}
\end{equation}

Thus when we include all directions  
\begin{equation}
\begin{split}
S = &  \frac{1}{\sqrt{\det (\eta+B)}} \prod_a 
(1 + \ha B_{ab } \Gamma^{a}\Gamma^{b}) \\
& = \frac{\sqrt{-\det g}}{\sqrt{-\det(g+B)}} \left\{1+
\frac{1}{2} B_{\mu\nu} \Gamma^{\mu\nu} + \frac{1}{2! 2^2} B_{\mu\nu}
B_{\rho \sigma} \Gamma^{\mu\nu\rho\sigma} + 
\frac{1}{3! 2^3} B_{\mu\nu} B_{\rho \sigma} B_{\tau \lambda}
\Gamma^{\mu\nu\rho \sigma\tau \lambda}
\right. \\ & \qquad \left.
+ \frac{1}{4! 2^4} B_{\mu\nu} B_{\rho \sigma} B_{\tau \lambda} B_{\alpha\beta}
\Gamma^{\mu\nu\rho \sigma\tau \lambda\alpha\beta} 
+ \frac{1}{5! 2^5} B_{\mu\nu} B_{\rho \sigma} B_{\tau \lambda} B_{\alpha\beta}
B_{\gamma\delta} \Gamma^{\mu\nu\rho \sigma\tau \lambda\alpha\beta
\gamma\delta}
\right\}
\\&= \frac{\sqrt{-\det g}}{\sqrt{-\det(g+B)}} \AE(\fs{B})
\end{split}
\end{equation}
where in the first line the product runs over all 2$\times$2 blocks, 
and the second line is obtained from expanding out the product in the
prior line and writing it in a covariant  form.
The last line follows from the explicit expansion of  $\AE(\fs{B})$.
(Of course many of the terms in the second line vanish trivially 
for $p<9$.)

Note that since $S$ generates a  Lorentz transformation, it satisfies 
$S S^{\transpose} = 1$, which implies an  interesting and useful identity
for $\AE$,
\begin{equation}
\AE(\fs{B}) \AE(-\fs{B}) = \frac{\det(g+B)}{\det g} \ .
\end{equation}

We finally note that equations~\eqref{mcondc} and~\eqref{mcondg} determine 
$M$ up to an arbitrary sign.

\section{Trace Formulas} \label{app:trace}

We wish to derive equation~\eqref{expgiveL}, and, in the process,
derive some other useful formulas.  The main such useful formula is
\begin{smaleq}
\begin{multline} \label{giveL}
\!\!\!
\Lambda^{\mu_1\dots\mu_q i_1\dots i_r}
= 32
\sum_{n,m}
    \frac{ q! (-1)^{\half r(r-1)} }{
    2^m m! (q-2m)! (n-r)! (p+1-n-q+r+2m)! 
    (\frac{p+1-n}{2}-\frac{q-r}{2}+m)!}
\\ \times
\frac{\theta^{[\mu_1\mu_2}}{2\pi\apr} \dots 
   \frac{\theta^{\mu_{2m-1}\mu_{2m}}}{2\pi\apr}
\epsilon^{\mu_{2m+1}\dots\mu_{q}]\sigma_1\dots\sigma_{n-r} 
   \tau_1\dots\tau_{p+1-n-q+r+2m}}
\\ \times
C^{(n)}_{i_1\dots i_r \sigma_1\dots\sigma_{n-r}}
\left[B^{\frac{p+1-n}{2}-\frac{q-r}{2}+m}\right]_{
    \tau_1\dots\tau_{p+1-n-q+r+2m}},
\end{multline}
\end{smaleq}%
for $q+r$ even (otherwise $\Lambda^{\mu_1\dots\mu_q i_1\dots i_r}$
vanishes, of course%
\iftoomuchdetail%
, because the trace of an odd number of $\Gamma$-matrices vanishes%
\fi%
).  We denote by
$\epsilon$
the antisymmetric volume element for the D$p$-brane.
Note that the sum is only over $n$ with the same parity as $p+1$; that
is over odd (even) $n$ for type IIA (IIB).

Once we have established equation~\eqref{giveL}---as we shortly will---then
obtaining~\eqref{expgiveL} is merely an
exercise in combinatorics.  Namely,
\begin{multline} \label{expandexpL}
\frac{\omega_{\mu_1\dots\mu_q}}{q!}\frac{\chi_{i_1\dots i_r}}{r!}
   \Lambda^{\mu_1\dots \mu_q i_1\dots i_r}
= 32\sum_{n,m} 
\frac{ \frac{1}{2^m m!} \frac{\theta^{\mu_1\mu_2}}{2\pi\apr} \dots 
   \frac{\theta^{\mu_{2m-1}\mu_{2m}}}{2\pi\apr}
\omega_{\mu_1\dots\mu_q}}{(q-2m)!}
\\ \times
\epsilon^{\mu_{2m+1}\cdots\mu_q \sigma_1 \cdots \sigma_{n-r} 
          \tau_1\cdots\tau_{p+1-n-q+r+2m}}
\left(\frac{(-1)^{\half r(r-1)}}{r!} \chi^{i_1\cdots i_r} 
   \frac{C^{(n)}_{i_1\cdots i_r \sigma_1\cdots \sigma_{n-r}}}{(n-r)!}
\right)
\\ \times
\frac{\left[B^{\frac{p+1-n}{2}-\frac{q-r}{2}+m}\right]_{
    \tau_1\dots\tau_{p+1-n-q+r+2m}}}{ (p+1-n-q+r+2m)! 
    (\frac{p+1-n}{2}-\frac{q-r}{2}+m)! }.
\end{multline}
It is not hard to see that this is an expanded version of
equation~\eqref{expgiveL}%
\putexpgiveL
Specifically, we can recognize
\begin{equation*}
\frac{ \frac{1}{2^m m!} \frac{\theta^{\mu_1\mu_2}}{2\pi\apr} \dots 
   \frac{\theta^{\mu_{2m-1}\mu_{2m}}}{2\pi\apr}
\omega_{\mu_1\dots\mu_q}}{(q-2m)!} 
dx^{\mu_{2m+1}}\wedge \cdots\wedge dx^{\mu_q}
\end{equation*}
as the $m^{\text{th}}$ term in the expansion of
$e^{-\co{\sfrac{\theta}{2\pi\apr}}} \omega$, where the minus sign
in the exponential comes from interchanging the order of the indices as per the
definition~\eqref{defco}.  Similarly, it is clear that
\begin{equation*}
\frac{(-1)^{\half r(r-1)}}{r!} \chi^{i_1\cdots i_r} 
   \frac{C^{(n)}_{i_1\cdots i_r \sigma_1\cdots \sigma_{n-r}}}{(n-r)!}
dx^{\sigma_1} \wedge \cdots \wedge dx^{\sigma_{n-r}}
= \co{\chi} C^{(n)}.
\end{equation*}
Finally, the Hodge dualization and the factor of $e^B$ is also
obvious.  The fact that the number of powers of $B$ in the expansion
of $e^B$ is related to the
number of powers of $\theta$ in $e^{\co{\sfrac{\theta}{2\pi\apr}}}$,
is due to the restriction to the
zero-form, since
$\frac{\omega_{\mu_1\dots\mu_q}}{q!}\frac{\chi_{i_1\dots i_r}}{r!}
   \Lambda^{\mu_1\dots \mu_q i_1\dots i_r}$
is a scalar.

Deriving equation~\eqref{giveL} is quite lengthy and tedious.
We start by considering the object
\begin{equation} \label{altL}
\tilde{\Lambda}_n^{\mu_1\cdots\mu_{q} i_1\cdots i_r} =
\Tr\left(\Gamma^{\mu_1\cdots\mu_q i_1\cdots i_r} \fs{C^{(n)}}
\AE(\fs{B}) \Gamma^0 \cdots \Gamma^p \right),
\end{equation}
and we observe that since the only transverse $\Gamma$ matrices, aside
from the $\Gamma^i$s, are attached to $C^{(n)}$, we must have
\begin{equation}
\tilde{\Lambda}_n^{\mu_1\cdots\mu_{q} i_1\cdots i_r} =
(-1)^{\frac{r(r-1)}{2}}
\Tr\left(\Gamma^{\mu_1\cdots\mu_q}
 \frac{C^{(n)}{^{i_1\cdots i_r}}_{\sigma_1\cdots\sigma_{n-r}}
\Gamma^{\sigma_1\cdots\sigma_{n-r}}}{(n-r)!}
\AE(\fs{B}) \Gamma^0 \cdots \Gamma^p \right).
\end{equation}
Thus, it is sufficient to consider
\begin{equation} \label{altLl}
\tilde{\Lambda}_{\tilde{n}}^{\mu_1\cdots\mu_q} = 
\frac{\epsilon_{\tau_0\cdots \tau_p}}{(p+1)!}
\Tr\left(\Gamma^{\mu_1\cdots\mu_q}
  \frac{C^{(\tilde{n})}_{\sigma_1\cdots\sigma_{\tilde{n}}}
\Gamma^{\sigma_1\cdots\sigma_{\tilde{n}}}}{n!}
\sum_m \frac{B^m_{\rho_1\cdots\rho_{2m}} \Gamma^{\rho_1\cdots\rho_{2m}}}%
            {m!(2m)!}
\Gamma^{\tau_0 \cdots \tau_p} \right).
\end{equation}
for arbitrary $\tilde{n},q$, with $\tilde{n}+q$ having the opposite
parity of $p$,
and where $B^m$ means
$\overbrace{B\wedge\cdots\wedge B}^{m\text{ factors}}$.
By replacing $\tilde{n}$ with $n-r$ and adding $r$ transverse indices to
$C^{(\tilde{n})}\rightarrow C^{(n)}$, 
where $r$ and $q$ have the same parity, we will recover
$(-1)^{\frac{r(r-1)}{2}}\tilde{\Lambda}_n^{\mu_1\cdots\mu_{q}i_1\cdots i_r}$,
where here $n$
and $p+1$ must also have the same parity.  In the following we will
drop the tilde on $\tilde{n}$.

We next observe that
\begin{equation}
\epsilon_{\tau_0\cdots \tau_p} \Gamma^{\tau_0 \cdots \tau_p}
\Gamma^{\mu_1\cdots\mu_q}
= (-1)^{\frac{q(q-1)}{2}}
  \frac{(p+1)!}{(p+1-q)!}
\epsilon_{\tau_0\cdots\tau_{p-q}}{^{\mu_1\cdots\mu_q}} 
\Gamma^{\tau_0\cdots \tau_{p-q}},
\end{equation}
where indices are raised and lowered with the closed string metric $g$.
\iftoomuchdetail%
The numerical factor comes from the $\left(\begin{smallmatrix} p+1 \\ q
\end{smallmatrix}\right)$ ways of matching the
$\mu$s with $\tau$, but $q!$ degeneracies therein.
\fi%
Therefore,%
\footnote{Here we clearly have $0\leq m \leq \frac{p+1}{2}$; however, in
practice some of these terms may vanish.
At any rate, the combinatorics will take care of
the region of summation, so we will never write it explicitly.}
\begin{equation} 
\tilde{\Lambda}_n^{\mu_1\cdots\mu_q} 
= (-1)^{\frac{q(q-1)}{2}}
\sum_m \frac{ \epsilon_{\tau_0\cdots \tau_{p-q}}{^{\mu_1\cdots\mu_q}}
   C^{(n)}_{\sigma_1\cdots\sigma_n}
   B^m_{\rho_1\cdots\rho_{2m}} }{n!m!(2m)!(p+1-q)!}
\Tr\left(\Gamma^{\sigma_1\cdots\sigma_n}
\Gamma^{\rho_1\cdots\rho_{2m}}
\Gamma^{\tau_0 \cdots \tau_{p-q}} \right).
\end{equation}
The usual trace rules tells us that, say, the first $s$ $\sigma$s should be
identified with the last $s$ $\tau$s and that the remaining $\sigma$s
should be identified with the first $n-s$ $\rho$s.  Identifying the
remaining $\rho$s with the remaining $\tau$s gives
$s=\frac{p+1+n-q}{2}-m$.
Including the combinatorics and signs 
\iftoomuchdetail%
(permutations are taken into account by the antisymmetric tensors out
front)
\fi%
gives
\begin{multline} \label{altLnoTr}
\tilde{\Lambda}_n^{\mu_1\cdots\mu_q} 
= 32 \sum_m (-1)^{\frac{n(n-1)}{2} + \frac{p(p+1)}{2} + pq
+ \frac{1}{2} \left(\frac{p+1+n-q}{2}-m\right)
   \left(\frac{p+1+n-q}{2}-m-1\right)}
\\ \times
\frac{\epsilon^{\tau_0\cdots \tau_{p-q}\mu_1\cdots\mu_q}
C^{(n)}_{\tau_{\frac{p+1-n-q}{2}+m}\cdots \tau_{p-q}
     \rho_1\cdots \rho_{\frac{n+q-p-1}{2}+m}}
\bigl(B^{m}\bigr)^{\rho_1\cdots \rho_{\frac{n+q-p-1}{2}+m} }{_{
        \tau_0\cdots \tau_{\frac{p+1-n-q}{2}+m-1}}}
}{m!\left(\frac{p+1+n-q}{2}-m\right)!
   \left(\frac{n+q-p-1}{2}+m\right)!\left(\frac{p+1-n-q}{2}+m\right)!}.
\end{multline}

Equation~\eqref{altLnoTr}, however, is not the most convenient form 
to work with; it would be more convenient for $B$ to never contract
with $C$.  Since every possible longitudinal index is attached to
$\epsilon$, it follows that $\rho_1\cdots \rho_{\frac{n+q-p-1}{2}+m}$
are equal to an appropriate subset of the $\mu$s.  Thus, we can
replace the $\rho$s with $\mu$s, and then replace the $\mu$s on
$\epsilon$ and $C^{(n)}$ with $\rho$s.  In that way, we have shifted some of
the $\mu$s from $\epsilon$ to $B$, and $C^{(n)}$ is fully contracted
with $\epsilon$.  Including permutations and combinatorics gives
\iftoomuchdetail
\begin{multline}
\tilde{\Lambda}_n^{\mu_1\cdots\mu_q} 
= 32 \sum_m \frac{q! 
(-1)^{
\frac{1}{2}\left(\frac{p+1+q-n}{2}-m\right)\left(\frac{p+1+q-n}{2}-m\right) 
+ \frac{1}{2}(p+q)(p+q+1)}
}{n!m!\left(\frac{n+q-p-1}{2}+m\right)!\left(\frac{p+1-n-q}{2}+m\right)!
\left(\frac{p+1-n+q}{2}-m\right)!}
\\ \times
\epsilon^{\sigma_1\cdots\sigma_n 
                \tau_0\cdots\tau_{\frac{p+1-n-q}{2}+m-1}
                [\mu_1\cdots \mu_{\frac{p+1+q-n}{2}-m}}
      C^{(n)}_{\sigma_1\cdots \sigma_n}
      \bigl(B^{m}\bigr)^{\mu_{\frac{p+1+q-n}{2}-m+1}\cdots\mu_q]}{_{
               \tau_0\cdots\tau_{\frac{p+1-n-q}{2}+m-1}}}.
\end{multline}
If we redefine $m\rightarrow \frac{p+1+q-n}{2}-m$, we get
\fi%
\begin{multline} \label{happyTr}
\tilde{\Lambda}_n^{\mu_1\cdots\mu_q}
= 32 \sum_m \frac{q! (-1)^{\frac{m(m+1)}{2} + \frac{(p+q)(p+q+1)}{2}}
}{n!m!(q-m)!(p+1-n-m)!(\frac{p+1+q-n}{2}-m)!}
\\ \times
\epsilon^{\sigma_1\cdots\sigma_n \tau_0\cdots \tau_{p-n-m} [\mu_1\cdots\mu_m}
C^{(n)}_{\sigma_1\cdots\sigma_n}
\bigl(B^{\frac{p+1+q-n}{2}-m}\bigr)^{\mu_{m+1}\cdots\mu_q]}{_{
   \tau_0\cdots\tau_{p-n-m}}}%
\iftoomuchdetail.\else,\fi
\end{multline}
\iftoomuchdetail\else%
where we have also redefined $m\rightarrow \frac{p+1+q-n}{2}-m$.
\fi%

Of course, what we really want is not $\tilde{\Lambda}$, but
\begin{smaleq}
\begin{multline} \iftoomuchdetail\else\label{me+o}\fi
\Lambda^{\mu_1\cdots \mu_q} = 
-\left(\frac{1}{g-B}g\right)^{\mu_1}{_{\nu_1}}
\cdots \left(\frac{1}{g-B}g\right)^{\mu_q}{_{\nu_q}}
\sum_n \tilde{\Lambda}_n^{\nu_1\cdots \nu_q} \\
= \left(\frac{1}{g-B}g\right)^{\mu_1}{_{\nu_1}}
\cdots \left(\frac{1}{g-B}g\right)^{\mu_q}{_{\nu_q}}
\sum_{n,m} \frac{32 q! (-1)^{\frac{m(m+1)}{2} + qp + \frac{p(p+1)}{2}+1}
}{n!m!(q-m)!(p+1-n-q+m)!(\frac{p+1-n-q}{2}+m)!}
\\ \times
\epsilon^{\sigma_1\cdots\sigma_n \tau_0\cdots \tau_{p-n-q+m} 
   [\nu_{m+1}\cdots\nu_q}
C^{(n)}_{\sigma_1\cdots\sigma_n}
\bigl(B^{\frac{p+1-n-q}{2}+m}\bigr)^{\nu_{1}\cdots\nu_m]}{_{
   \tau_0\cdots\tau_{p-n-q+m}}},
\end{multline}
\end{smaleq}%
where we have replaced $m\rightarrow q-m$
\iftoomuchdetail%
(which makes sense since the combinatorics requires
$0\leq m\leq q$)
\fi%
and rearranged the $\nu$ indices.
To evaluate this, it is convenient to distinguish between an even and
odd number of $\mu$s attached to $B^{\cdot\cdot}$; i.e. $m$ even or odd.
\iftoomuchdetail%
So we write,
\begin{smaleq}
\begin{multline} \label{me+o}
\Lambda^{\mu_1\cdots \mu_q}
= 32 \left(\frac{1}{g-B}g\right)^{\mu_1}{_{\nu_1}}
\cdots \left(\frac{1}{g-B}g\right)^{\mu_q}{_{\nu_q}}
\\ \times
\sum_{n,m}
\biggl\{
\frac{q! (-1)^{m+pq+ \frac{p(p+1)}{2}}}{n!(2m)!(q-2m)!(p+1-n-q+2m)!
   (\frac{p+1-n-q}{2}+2m)!}
\\ \times
\epsilon^{\sigma_1\cdots\sigma_n \tau_0\cdots \tau_{p-n-q+2m} 
   [\nu_{2m+1}\cdots\nu_q}
C^{(n)}_{\sigma_1\cdots\sigma_n}
\bigl(B^{\frac{p+1-n-q}{2}+2m}\bigr)^{\nu_{1}\cdots\nu_{2m}]}{_{
   \tau_0\cdots\tau_{p-n-q+2m}}}
\\ -
\frac{q! (-1)^{m+pq +\frac{p(p+1)}{2}}}{n!(2m+1)!(q-2m-1)!(p+2-n-q+2m)!
   (\frac{p+3-n-q}{2}+2m)!}
\\ \times
\epsilon^{\sigma_1\cdots\sigma_n \tau_0\cdots \tau_{p+1-n-q+2m} 
   [\nu_{2m+2}\cdots\nu_q}
C^{(n)}_{\sigma_1\cdots\sigma_n}
\bigl(B^{\frac{p+3-n-q}{2}+2m}\bigr)^{\nu_{1}\cdots\nu_{2m+1}]}{_{
   \tau_0\cdots\tau_{p+1-n-q+2m}}}
\biggl\}.
\end{multline}
\end{smaleq}%
\fi
The point is that
\begin{subequations} \label{wbasb}
\begin{multline}
\bigl(B^{\frac{p+1-n-q}{2}+2m}\bigr)^{\nu_{1}\cdots\nu_{2m}}{_{
   \tau_0\cdots\tau_{p-n-q+2m}}}
\\= \sum_s \frac{(-1)^s (p+1-n-q+2m)! (2m)! 
\left(\frac{p+1-n-q}{2}+2m\right)!
}{2^{\frac{p+1-n-q}{2}+2m-2s}\left(\frac{p+1-n-q}{2}+m-s\right)! (2s)! (m-s)!}
B^{[\nu_1\nu_2} \cdots B^{\nu_{2m-2s-1}\nu_{2m-2s}}
\\ \times
B^{\nu_{2m-2s+1}}{_{[\tau_0}} \cdots B^{\nu_{2m}]}{_{\tau_{2s-1}}}
B_{\tau_{2s}\tau_{2s+1}} \cdots B_{\tau_{p-n-q+2m-1}\tau_{p-n-q+2m}]},
\end{multline}
and
\begin{multline}
\bigl(B^{\frac{p+3-n-q}{2}+2m}\bigr)^{\nu_{1}\cdots\nu_{2m+1}}{_{
   \tau_0\cdots\tau_{p+1-n-q+2m}}}
\\ = \sum_s \frac{(-1)^s (p+2-n-q+2m)! (2m+1)!
   \left(\frac{p+3-n-q}{2}+2m\right)!
}{2^{\frac{p+1-n-q}{2}+2m-2s}
  \left(\frac{p+1-n-q}{2}+m-s\right)! (2s+1)! (m-s)!}
B^{[\nu_1\nu_2} \cdots B^{\nu_{2m-2s-1}\nu_{2m-2s}} 
\\ \times
B^{\nu_{2m-2s+1}}{_{[\tau_0}} \cdots B^{\nu_{2m+1}]}{_{\tau_{2s}}}
B_{\tau_{2s+1}\tau_{2s+2}} \cdots B_{\tau_{p-n-q+2m}\tau_{p+1-n-q+2m}]}.
\end{multline}
\end{subequations}
So 
substituting the
identities~\eqref{wbasb} into~\eqref{me+o}, gives%
\footnote{We drop an overall sign of $(-1)^{\frac{p(p+1)}{2}+1}$; this is
of no physical consequence.}
\begin{smaleq}
\begin{multline} \label{lwob}
\Lambda^{\mu_1\cdots\mu_q}
= 32 \left(\frac{1}{g-B}g\right)^{\mu_1}{_{[\nu_1}}
\cdots \left(\frac{1}{g-B}g\right)^{\mu_q}{_{\nu_q]}}  
C^{(n)}_{\sigma_1\cdots\sigma_n}
\\ \times
\sum_{n,m,s} \biggl\{
\frac{q!(-1)^{m+s+pq}}{n!(q-2m)!
2^{\frac{p+1-n-q}{2}+2m-2s}\left(\frac{p+1-n-q}{2}+m-s\right)! (2s)!(m-s)!}
\\ \times
\epsilon^{\sigma_1\cdots\sigma_n \tau_0\cdots \tau_{p-n-q+2m}}{_{
   \tau_{p-n-q+2m+1} \tau_{p-n}}}
B^{\nu_1\nu_2} \cdots B^{\nu_{2m-2s-1}\nu_{2m-2s}}
B^{\nu_{2m-2s+1}\tau_0} \cdots B^{\nu_{2m} \tau_{2s-1}}
\\ \times
B^{\tau_{2s}\tau_{2s+1}} \cdots B^{\tau_{p-n-q+2m-1}\tau_{p-n-q+2m}}
g^{\nu_{2m+1}\tau_{p-n-q+2m+1}} \cdots g^{\nu_q \tau_{p-n}}
\\
-\frac{q!(-1)^{m+s+pq}}{n!(q-2m-1)!
2^{\frac{p+1-n-q}{2}+2m-2s}\left(\frac{p+1-n-q}{2}+m-s\right)! 
(2s+1)!(m-s)!}
\\ \times
\epsilon^{\sigma_1\cdots\sigma_n \tau_0\cdots \tau_{p+1-n-q+2m}}{_{
   \tau_{p+2-n-q+2m} \tau_{p-n}}}
B^{\nu_1\nu_2} \cdots B^{\nu_{2m-2s-1}\nu_{2m-2s}}
B^{\nu_{2m-2s+1}\tau_0} \cdots B^{\nu_{2m+1} \tau_{2s}}
\\ \times
B^{\tau_{2s+1}\tau_{2s+2}} \cdots B^{\tau_{p-n-q+2m}\tau_{p+1-n-q+2m}}
g^{\nu_{2m+2}\tau_{p+2-n-q+2m}} \cdots g^{\nu_q \tau_{p-n}}
\biggr\}.
\end{multline}
\end{smaleq}%

Now we set $m=t+s$, and also cyclically permute the $\tau$-indices
between the last
set of $B$s and the
$g$s, to rewrite equation~\eqref{lwob} as
\begin{smaleq}
\begin{multline}
\Lambda^{\mu_1\cdots\mu_q}
= 32 \left(\frac{1}{g-B}\right)^{\mu_1[\nu_1}
\cdots \left(\frac{1}{g-B}\right)^{|\mu_q|\nu_q]}
C^{(n)}_{\sigma_1\cdots\sigma_n}
\\ \times
\sum_{n,s,t} \biggl\{
\frac{q!(-1)^{t+pq}}{n!(q-2t-2s)!
2^{\frac{p+1-n-q}{2}+2t}\left(\frac{p+1-n-q}{2}+t\right)! (2s)!t!}
\\ \times
\epsilon^{\sigma_1\cdots\sigma_n\tau_0\cdots\tau_{p-n}}
B_{\nu_1\nu_2} \cdots B_{\nu_{2t-1}\nu_{2t}}
B_{\nu_{2t+1}\tau_0} \cdots B_{\nu_{2t+2s} \tau_{2s-1}}
\\ \times
g_{\nu_{2s+2t+1}\tau_{2s}} \cdots g_{\nu_q \tau_{q-2t-1}}
B_{\tau_{q-2t}\tau_{q-2t+1}} \cdots B_{\tau_{p-n-1}\tau_{p-n}}
\\
-\frac{q!(-1)^{t+pq}}{n!(q-2t-2s-1)!
2^{\frac{p+1-n-q}{2}+2t}\left(\frac{p+1-n-q}{2}+t\right)! 
(2s+1)!t!}
\\ \times
\epsilon^{\sigma_1\cdots\sigma_n \tau_0\cdots\tau_{p-n}}
B_{\nu_1\nu_2} \cdots B_{\nu_{2t-1}\nu_{2t}}
B_{\nu_{2t+1}\tau_0} \cdots B_{\nu_{2t+2s+1} \tau_{2s}}
\\ \times
g_{\nu_{2s+2t+2}\tau_{2s+1}} \cdots g_{\nu_q \tau_{q-2t-1}}
B_{\tau_{q-2t}\tau_{q-2t+1}} \cdots B_{\tau_{p-n-1}\tau_{p-n}}
\biggl \}.
\end{multline}
\end{smaleq}%
This can immediately be recombined into
\begin{smaleq}
\begin{multline} \label{almostL}
\Lambda^{\mu_1\cdots\mu_q}
= 32 \sum_{n,t} \frac{\theta^{[\mu_1\mu_2}}{2\pi\apr}
\cdots \frac{\theta^{\mu_{2t-1}\mu_{2t}}}{2\pi\apr}
\left(\frac{1}{g-B}\right)^{\mu_{2t+1}|\nu_{2t+1}|} \cdots 
\left(\frac{1}{g-B}\right)^{\mu_q]\nu_q}
\\ \times
\frac{q!(-1)^{pq}}{n!
2^{\frac{p+1-n-q}{2}+2t}\left(\frac{p+1-n-q}{2}+t\right)!t!}
\epsilon^{\sigma_1\cdots\sigma_n\tau_0\cdots\tau_{p-n}}
C^{(n)}_{\sigma_1\cdots\sigma_n}
B_{\tau_{q-2t}\tau_{q-2t+1}} \cdots B_{\tau_{p-n-1}\tau_{p-n}}
\\ \times
\sum_{s} \frac{(-1)^s 
B_{\nu_{2t+1}\tau_0} \cdots B_{\nu_{2t+s} \tau_{s-1}}
g_{\nu_{2t+s+1}\tau_{s}} \cdots g_{\nu_q \tau_{q-2t-1}}
}{(q-2t-s)! s!}
\\
= 32 \sum_{n,m} \frac{q!}{n!
2^{\frac{p+1-n-q}{2}+2m}(q-2m)!\left(\frac{p+1-n-q}{2}+m\right)!m!}
\frac{\theta^{[\mu_1\mu_2}}{2\pi\apr}
\cdots \frac{\theta^{\mu_{2m-1}\mu_{2m}}}{2\pi\apr}
\\ \times
\epsilon^{\mu_{2m+1}\cdots\mu_q]\sigma_1\cdots\sigma_n
  \tau_0\cdots\tau_{p-n-q+2m}}
C^{(n)}_{\sigma_1\cdots\sigma_n}
B_{\tau_0\tau_1} \cdots B_{\tau_{p-n-q+2m-1}\tau_{p-n-q+2m}}
\end{multline}
\end{smaleq}%
where we have noted that
$\frac{1}{g-B}B\frac{1}{g+B} = -\frac{\theta}{2\pi\apr}$, and recognized the
binomial theorem in the second step.
\iftoomuchdetail%
Also, in the last step, we have noted that $(n+p)q = (p+1+n+q)q-(q+1)q
= 0 \bmod 2$.%
\fi%

Replacing $n$ with $n-r$ and adding in the transverse indices plus
the extra sign, $(-1)^{\half r(r-1)}$, as discussed surrounding
equation~\eqref{altLl}, turns equation~\eqref{almostL} into
equation~\eqref{giveL}.  This is the desired result.

\section{Integrals For the Two Open String Amplitude} \label{sec:ints}

Here we list the exact formulas for the integrals~\eqref{ints}.
Recall that $y = -\cot(\pi \tau)$.
\begin{subequations} \label{fullints}
\begin{smaleq}
\begin{gather}
\begin{split} \label{fullint0} \raisetag{2\baselineskip}
\int_{-\infty}^\infty dy \; \frac{A_2 (y)}{1+ y^2} 
&= \pi i 2^{2 \apr k_1 \cdot k_2} \,
\int_{0}^1 d\tau \abs{\cos \pi\tau}^{2\apr k_1\cdot k_2}
\exp\left\{\frac{i}{2} (k_1\times k_2) 
    \left[1-2\tau-\epsilon(\thalf-\tau)\right]\right\} \\
&=  \pi i 2^{2 \apr k_1 \cdot k_2+1} \, 
\int_{0}^\half d\tau \cos^{2\apr k_1\cdot k_2} \pi \tau \,
\cos\left[(k_1\times k_2) \tau\right]
\\
&= \pi i 
   \frac{\Gamma(2\apr k_1\cdot k_2 + 1)}%
        {\Gamma\left(\apr k_1\cdot k_2+1+\frac{k_1\times k_2}{2\pi}\right)
         \Gamma\left(\apr k_1\cdot k_2+1-\frac{k_1\times k_2}{2\pi}\right)}, 
\end{split}
\end{gather}\end{smaleq}\begin{gather}
\begin{split} \label{fullintcot}
\int_{-\infty}^\infty dy \; &\frac{A_2 (y)}{y(1+ y^2)} 
= -
   \frac{(k_1\times k_2) \; \Gamma(2\apr k_1\cdot k_2)}%
        {\Gamma\left(\apr k_1\cdot k_2+1+\frac{k_1\times k_2}{2\pi}\right)
         \Gamma\left(\apr k_1\cdot k_2+1-\frac{k_1\times k_2}{2\pi}\right)},
\end{split}
\end{gather}\begin{gather}
\begin{split} \label{fullintcsc2}
(1+\apr t)\int_{-\infty}^\infty dy \; \frac{A_2 (y)}{y^2} 
&= -\pi i
   \frac{4\Gamma(2\apr k_1\cdot k_2)}%
        {\Gamma\left(\apr k_1\cdot k_2+\frac{k_1\times k_2}{2\pi}\right)
         \Gamma\left(\apr k_1\cdot k_2-\frac{k_1\times k_2}{2\pi}\right)}.
\end{split}
\end{gather}
\end{subequations}
On the right-hand sides of equations~\eqref{fullints}, we have
suppressed the factor of $(2\pi)^{p+1} \delta(k_1+k_2+q_\parallel)$.
We obtain equations~\eqref{ints} by taking $\apr k_1\cdot k_2\rightarrow 0$ and
using the identity
\begin{equation} \label{lowenergy}
\frac{\Gamma(2\apr k_1\cdot k_2 + 1)}%
     {\Gamma\left(\apr k_1\cdot k_2+1+\frac{k_1\times k_2}{2\pi}\right)
      \Gamma\left(\apr k_1\cdot k_2+1-\frac{k_1\times k_2}{2\pi}\right)}
= \frac{\sin \frac{k_1\times k_2}{2}}{\frac{k_1\times k_2}{2}}
+\order{\apr}.
\end{equation}
We recognize the right-hand side as the $\ast_2$-operation.\cite{garousi,lm2}.

\section{Comparison of the Amplitude to the Action} \label{sec:verify}

In this appendix, we compare the action~\eqref{rrf} to the
amplitude~\eqref{totamp}.  In section~\ref{sec:expamp}, we rewrite the
amplitude~\eqref{totamp} in such a way as to make the first few terms
in an expansion of~\eqref{rrf} manifest.  For the reader's
edification, we explicitly expand the
action~\eqref{rrf}, in section~\ref{sec:exps}, to quadratic order in
the open string modes.

\subsection{The Explicit Form of the Amplitude} \label{sec:expamp}

The amplitude we computed in section~\ref{sec:amp} corresponds to the
action~\eqref{amp0}, which we write explicitly as
\begin{multline} \label{s0}
S =  \frac{\lambda}{2} \kappa_{10} \mu_{p} \int \sqrt{-\det g} 
\Tr \biggl[ \Lambda
+ 2\pi \apr \frac{f_{\mu\nu}}{2!} \Lambda^{\mu\nu}
+ 2\pi\apr D_\mu \phi_i \Lambda^{\mu i}
+ 2 \pi \apr i \com{\phi_i}{\phi_j}_\ast \Lambda^{ij}\\
+ \frac{1}{2} (2\pi \apr)^2 \frac{(f\wedge f)_{\mu \nu\rho \sigma}}{4!}
  \Lambda^{\mu \nu \rho \sigma}
+ (2\pi \apr)^2 \frac{f_{\mu \nu}}{2!} D_{\rho}\phi_i
  \Lambda^{\mu \nu \rho i}
+ \frac{1}{2} (2\pi \apr)^2 D_{\mu} \phi_i D_{\nu} \phi_j
  \Lambda^{\mu i \nu j}
\biggr],
\end{multline}
where it is understood that we should only keep terms up to quadratic
in open string modes.
Recalling equation~\eqref{expgiveL},
\putexpgiveL
we can immediately rewrite equation~\eqref{s0} as,
\begin{smaleq}
\begin{multline} \label{sl}
S = 
\kappa_{10} \mu_{p} \Tr \Biggl\{
\biggl(
\int \left[e^{-\co{\sfrac{\theta}{2\pi\apr}}} \left( 1
+ 2\pi \apr f
+ \frac{1}{2} (2\pi \apr f)^2 \right) \right] C e^B \\
+ \int
   \co{2 \pi \apr i \com{\phi_i}{\phi_j}_\ast} C e^B \biggr)\\
+ \frac{\lambda}{2}
 \int \sqrt{-\det g} \left[ 2\pi\apr D_\mu \phi_i \Lambda^{\mu i}
+ (2\pi \apr)^2 \frac{f_{\mu \nu}}{2!} D_{\rho}\phi_i
  \Lambda^{\mu \nu \rho i}
+ \frac{1}{2} (2\pi \apr)^2 D_{\mu} \phi_i D_{\nu} \phi_j
  \Lambda^{\mu i \nu j} \right]
\Biggr\}\iftoomuchdetail,\else.\fi
\end{multline}
\end{smaleq}%
\iftoomuchdetail%
where we have used the notational simplification $\int \sqrt{-\det g}
\left[\star \omega\right]_{\text{0-form}} = \int \omega$.
\fi%

All but the last line of equation~\eqref{sl} is clearly the quadratic
expansion of
\begin{equation}
S = \kappa_{10} \mu_p \Tr \int
\left(e^{-\co{\sfrac{\theta}{2\pi\apr}}} e^{-2\pi\apr f} \right)
e^{-\co{2\pi\apr i \com{\phi}{\phi}}} C e^B.
\end{equation}
The second line of equation~\eqref{sl} will contain the pullback
terms; however, this is somewhat more difficult to see
because it involves objects which have both longitudinal and
transverse indices, while the identity~\eqref{expgiveL} separates
those indices.  Nevertheless, we can use
the identity~\eqref{expgiveL} if we include the indices explicitly 
(cf.\ equation~\eqref{giveL}).
For the first term of the second line of equation~\eqref{sl}, we find
\begin{equation} \label{firstpb}
\frac{\lambda}{2} \int \sqrt{-\det g} 
\left[2\pi\apr D_\mu \phi_i \Lambda^{\mu i}\right]
= \int \left[\frac{D_\mu X^i
\bigl(Ce^B\bigr)_{i\nu_1\cdots\nu_{p}}}{p!}
dx^\mu dx^{\nu_1} \cdots dx^{\nu_{p}} \right].
\end{equation}
We indeed recognize equation~\eqref{firstpb} as the first term in the
pullback $\Pb Ce^B$; see equation~\eqref{defpb}.
The second term similarly gives
\begin{multline} \label{fpb}
\frac{\lambda}{2}
\int \sqrt{-\det g} \left[ (2\pi \apr)^2 \frac{f_{\mu \nu}}{2!} D_{\rho}\phi_i
  \Lambda^{\mu \nu \rho i} \right]
\\=
\int \left[ \frac{2\pi\apr f_{\mu\nu}(D_{\rho} X^i)
\bigl(Ce^B\bigr)_{i\sigma_1\cdots\sigma_{p-2}}}{2!(p-2)!}
dx^\mu dx^\nu dx^\rho dx^{\sigma_1} \cdots dx^{\sigma_{p-2}} 
\right. \\ \left.
+ \frac{3!}{2} \frac{\theta^{\mu\nu}}{2\pi\apr} 
  \frac{(2\pi\apr) f_{[\mu\nu} (D_{\rho]} X^i)
\bigl(Ce^B\bigr)_{i\sigma_1\cdots\sigma_{p}}}{2!p!}
dx^\rho dx^{\sigma_1} \cdots dx^{\sigma_{p}} 
\right],
\end{multline}
Note not only the linear contribution to the pullback, with and
without $f$, but also the
contribution from $e^{-\co{\sfrac{\theta}{2\pi\apr}}}$ on the pullback
(and $F$).  Furthermore, equations~\eqref{firstpb} and~\eqref{fpb} contain
the only terms linear in the pullback and no more than quadratic in
open string modes.

The final term is similar; it gives
\begin{multline} \label{pb2}
\frac{\lambda}{2}
\int \sqrt{-\det g} \frac{1}{2} (2\pi \apr)^2 D_{\mu} \phi_i D_{\nu} \phi_j
  \Lambda^{\mu i \nu j}
\\ =
\int \left[ \frac{D_{\mu}X^i D_{\nu} X^j 
\bigl(Ce^B\bigr)_{ij\sigma_1\cdots\sigma_{p-1}}}{2!(p-1)!}
dx^\mu dx^\nu dx^{\sigma_1} \cdots dx^{\sigma_{p-1}} 
\right. \\ \left.
+ \frac{1}{2} \frac{\theta^{\mu\nu}}{2\pi\apr}
\frac{D_{\mu}X^i D_{\nu} X^j
\bigl(Ce^B\bigr)_{ij\sigma_1\cdots\sigma_{p+1}}}{2!(p+1)!}
dx^\mu dx^\nu dx^{\sigma_1} \cdots dx^{\sigma_{p+1}} \right],
\end{multline}
and we see the quadratic approximation to the pullback, along with the
relevant part of $e^{-\co{\sfrac{\theta}{2\pi\apr}}}$,
as in equation~\eqref{rrf}.
Thus, indeed, equation~\eqref{amp0} is the quadratic approximation to
equation~\eqref{rrf}, as claimed.

\subsection{The Expansion of the Proposed Action} \label{sec:exps}

\iftoomuchdetail
Explicitly, if slightly formally, we can rewrite the
action~\eqref{rrf} as
\begin{multline} \label{monsterc}
\int \left(e^{-\co{\sfrac{\theta}{2\pi\apr}}} e^{2\pi\apr f} \Pb\right) 
e^{-2\pi\apr i \co{\phi} \co{\phi}} C e^B
= \int
\sum_{k,j,l,q,n,m}
\frac{(-2\pi\apr i)^k}{k!} 
\phi^{i_1} \dots \phi^{i_{2k}} \frac{\p}{\p x^{i_1}} \dots
  \frac{\p}{\p x^{i_{2k}}}
\\ \times
\left[
\frac{(-1)^j}{j! 2^j}
\frac{\theta^{\mu_1\mu_2}}{2\pi\apr}\dots
   \frac{\theta^{\mu_{2j-1}\mu_{2j}}}{2\pi\apr} 
  \frac{\p}{\p x^{\mu_1}} \dots \frac{\p}{\p x^{\mu_{2j}}}
\frac{(2\pi\apr)^l}{l! 2^l} 
f_{\nu_1 \nu_2} \dots f_{\nu_{2l-1}\nu_{2l}} dx^{\nu_1}
  \dots dx^{\nu_{2l}}
\right.\\\left. \times
\frac{(2\pi\apr)^q(n-2k)!}{(n-2k-q)! q!} 
  D_{\rho_1} \phi^{j_1} \dots D_{\rho_q} \phi^{j_q}
  dx^{\rho_1} \dots dx^{\rho_q} \right]
\\ \times
\frac{1}{(2k)!(n-2k)!} C^{(n)}_{k_1\dots k_{2k}j_1\dots j_q \sigma_1\dots
  \sigma_{n-2k-q}} dx^{k_1}\dots dx^{k_{2k}} dx^{\sigma_1} \dots 
  dx^{\sigma_{n-2k-q}}
\\ \times
\frac{1}{m! 2^m} B_{\tau_1\tau_2}\dots B_{\tau_{2m-1}\tau_{2m}} dx^{\tau_1}
  \dots dx^{\tau_{2m}}.
\end{multline}
The integration picks out the $(p+1)$-form and so forces $n+2(m+l-j-k) = p+1$.
Also, we require, of course, that all of the $\frac{\p}{\p x}$s
``eat up'' some $dx$.  The square brackets prohibit a $\frac{\p}{\p x}$
inside the brackets from ``eating'' an outside $dx$ (and vice versa).
These requirements imply some restrictions---namely $2k+q\leq n$ (so
that there are not more
transverse indices on $C^{(n)}$ than there are indices) and $2j\leq 2l+q$
(so that $\theta$ doesn't contract with more indices than
exist)---which are
automatically enforced by the combinatorical factors.
\iftoomuchdetail
Most of these numerical factors should be self-evident.  For example, $C^{(n)}$
comes in with a $\frac{1}{(2k)!(n-2k)!}$ because of the ostensibly
$2k$ transverse
indices and (ostensibly) $n-2k$ longitudinal indices.
The combinatorical factor is not $1/n!$, because
we are distinguishing between transverse and longitudinal indices.  
It also is not $1/(2k+q)!~1/(n-2k-q)!$ for transverse and longitudinal
indices respectively, because the difference, $q$, in the number of
longitudinal and transverse indices is effected by the pullback.  (Said
another way, $\Pb(C^{(n)})$ is a $[2k+(n-2k)]$-form, not a
$[(2k+q)+(n-2k-q)]$-form.)
The numerical factor for the pullback 
is associated with the $DX$s, and is
combinatorical.
\fi%
When we write~\eqref{monsterc} in terms of indices rather than as a form,
we will get an additional factor of
\begin{equation} \label{en}
(-1)^{k} (2k)! \times (-1)^j \frac{(2l+q)!}{(2l+q-2j)!}.
\end{equation}
\iftoomuchdetail
The first factor is the number of ways of
contracting $\p/(\p x^{i})$s with $dx^k$s and the second factor is the
combinatorics for the contraction of $\p/(\p x^\mu)$s with
$dx^\nu$s and/or $dx^\rho$s.  The signs come from undoing the ``last to
first'' order of contractions.
\fi
\fi

We can 
\iftoomuchdetail%
now
\fi%
explicitly write the action~\eqref{rrf} term by term.  We find,
\begin{smaleq}
\begin{align} \label{fmc}
&\int \left(e^{-\co{\sfrac{\theta}{2\pi\apr}}}
e^{2\pi\apr f} \Pb\right) e^{-2\pi \apr i \co{\com{\phi}{\phi}}} C e^B
= \int d^{p+1} x \sqrt{-\det g} \; \epsilon^{\mu_1\dots\mu_{p+1}}
\nonumber \\* & \times \Biggl\{ 
\sum_n \frac{1}{n!\, (\frac{p+1-n}{2})!\, 2^{\frac{p+1-n}{2}}}
  C^{(n)}_{\mu_1\dots\mu_n} B_{\mu_{n+1}\mu_{n+2}} \dots
  B_{\mu_p\mu_{p+1}}
\nonumber \\ &
+ \sum_n \frac{2\pi\apr}{2!\,n!\,(\frac{p+1-n}{2}-1)!\,2^{\frac{p+1-n}{2}-1}}
  f_{\mu_1\mu_2} C^{(n)}_{\mu_3\dots\mu_{n+2}}
  B_{\mu_{n+3}\mu_{n+4}} \dots B_{\mu_p\mu_{p+1}}
\nonumber \\ &
+ \sum_n \frac{1}{2!\, n!\, (\frac{p+1-n}{2})!\, 2^{\frac{p+1-n}{2}}}
  \left(\Theta^{\nu_1\nu_2} f_{\nu_1\nu_2}\right)
  C^{(n)}_{\mu_1\dots\mu_n} B_{\mu_{n+1}\mu_{n+2}} \dots
  B_{\mu_p\mu_{p+1}}
\nonumber \\ &
+ \sum_n \frac{2\pi\apr}{(n-1)!\, (\frac{p+1-n}{2})!\, 2^{\frac{p+1-n}{2}}}
  D_{\mu_1} \phi^i
  C^{(n)}_{i\mu_2\dots\mu_n} B_{\mu_{n+1}\mu_{n+2}} \dots
  B_{\mu_p\mu_{p+1}}
\nonumber \\ &
+ \sum_n \frac{(2\pi\apr)^2}%
     {2^3\, n!\, (\frac{p+1-n}{2}-2)!\, 2^{\frac{p+1-n}{2}-2}}
  f_{\mu_1\mu_2} f_{\mu_3\mu_4} C^{(n)}_{\mu_5\dots\mu_{n+4}}
  B_{\mu_{n+5}\mu_{n+6}} \dots B_{\mu_p\mu_{p+1}}
\nonumber \\ &
+ \sum_n \frac{2\pi\apr\; 4!}%
         {2^5\, n!\, (\frac{p+1-n}{2}-1)!\, 2^{\frac{p+1-n}{2}-1}}
  \theta^{\nu_1\nu_2} f_{[\nu_1\nu_2}f_{\mu_1\mu_2]} 
  C^{(n)}_{\mu_3\dots\mu_{n+2}}
  B_{\mu_{n+3}\mu_{n+4}} \dots B_{\mu_p\mu_{p+1}}
\nonumber \\ &
+ \sum_n \frac{4!}%
         {2^6\,n!\,(\frac{p+1-n}{2})!\, 2^{\frac{p+1-n}{2}}}
  \theta^{\nu_1\nu_2} \theta^{\nu_3\nu_4} f_{[\nu_1\nu_2}f_{\nu_3\nu_4]} 
  C^{(n)}_{\mu_1\dots\mu_{n}}
  B_{\mu_{n+1}\mu_{n+2}} \dots B_{\mu_p\mu_{p+1}}
\nonumber \\ &
+ \sum_n \frac{(2\pi\apr)^2}%
         {2\,(n-1)!\,(\frac{p+1-n}{2}-1)!\, 2^{\frac{p+1-n}{2}-1}}
  f_{\mu_1\mu_2} D_{\mu_3} \phi^i
  C^{(n)}_{i\mu_4\dots\mu_{n+2}}
  B_{\mu_{n+3}\mu_{n+4}} \dots B_{\mu_p\mu_{p+1}}
\nonumber \\ &
+ \sum_n \frac{2\pi\apr\, 3!}%
         {2^2\,(n-1)!\,(\frac{p+1-n}{2})!\, 2^{\frac{p+1-n}{2}}}
  \theta^{\nu_1\nu_2}f_{[\nu_1\nu_2} D_{\mu_1]} \phi^i
  C^{(n)}_{i\mu_2\dots\mu_{n}}
  B_{\mu_{n+1}\mu_{n+2}} \dots B_{\mu_p\mu_{p+1}}
\nonumber \\ &
+ \sum_n \frac{(2\pi\apr)^2}%
         {2!\,(n-2)!\,(\frac{p+1-n}{2})!\, 2^{\frac{p+1-n}{2}}}
  D_{\mu_1} \phi^{i_1} D_{\mu_2} \phi^{i_2} C^{(n)}_{i_1i_2\mu_3\dots\mu_{n}}
  B_{\mu_{n+1}\mu_{n+2}} \dots B_{\mu_p\mu_{p+1}}
\nonumber \\ &
+ \sum_n \frac{2\pi\apr}%
         {2!\,(n-2)!\,(\frac{p+1-n}{2}+1)!\, 2^{\frac{p+1-n}{2}+1}}
  \theta^{\nu_1\nu_2} D_{\nu_1} \phi^{i_1} D_{\nu_2} \phi^{i_2} 
  C^{(n)}_{i_1i_2\mu_1\dots\mu_{n-2}}
  B_{\mu_{n-1}\mu_{n}} \dots B_{\mu_p\mu_{p+1}}
\nonumber \\ &
+ \sum_n \frac{2\pi\apr i}%
         {(n-2)!\,(\frac{p+1-n}{2}+1)!\, 2^{\frac{p+1-n}{2}+1}}
  \phi^{i_1} \phi^{i_2} C^{(n)}_{i_1i_2\mu_1\dots\mu_{n-2}}
  B_{\mu_{n-1}\mu_{n}} \dots B_{\mu_p\mu_{p+1}}  \Biggr\}
\nonumber \\* &
+ \dots,
\end{align}
\end{smaleq}%
where the omitted terms involve at least three open strings.
Note that the explicit metric dependence cancels the metric dependence
of the $\epsilon$-tensor.


\begin{thebibliography}{99}
 

\bibitem{miao} M. Li, \ct{Boundary States of D-Branes and Dy-Strings}
\npb{460}{1996}{351--361}; \phepth{9510161}.

\bibitem{ghm} M.~Green, J.~A.~Harvey, G.~Moore, \ct{I-Brane Inflow and
Anomalous Couplings on D-Branes} \cqg{14}{1997}{47--52}; \phepth{9605033}.

\bibitem{bwbw} E.~Witten, \ct{Small Instantons in String Theory},
\npb{460}{1996}{541--559}; \phepth{9511030}.

\bibitem{bwbd} M.~R.~Douglas, \ct{Branes within Branes} in
\bt{Strings, Branes and Dualities, Proceedings of the Carg\`{e}se '97
NATO ASI} (L. Baulieu, {\em et al.\/} ed.) Kluwer Press,
1998;
\phepth{9512077}.

\bibitem{mm} R.~Minasian and G.~Moore, \ct{K-theory and Ramond-Ramond
charge} \jhep{11}{1997}{002}; \phepth{9710230}.

\bibitem{witk} E.~Witten, \ct{D-Branes and K-Theory}
\jhep{12}{1998}{019}; \phepth{9810188}.
  
\bibitem{myers} R.~C.~Myers, \ct{Dielectric Branes}
\jhep{12}{1999}{022}; \phepth{9910053}.

\bibitem{taylor} W.~Taylor and M.~Van Raamsdonk, \ct{Multiple
$D0$-branes in Weakly Curved Backgrounds} \npb{558}{1999}{63--95},
\phepth{9904095}; \\
W.~Taylor and M.~Van Raamsdonk,
\ct{Multiple $Dp$-branes in Weak Background Fields}
\npb{573}{2000}{703--734}, \phepth{9910052}.

\bibitem{cds} A.~Connes, M.~R.~Douglas and A.~Schwarz,
\ct{Noncommutative geometry and matrix theory: Compactification on tori}
\jhep{02}{1998}{003}; \phepth{9711162}.

\bibitem{dh} M.~R.~Douglas and C.~Hull, \ct{D-branes and the Noncommutative
Torus} \jhep{02}{1998}{008}; \phepth{9711165}.

\bibitem{sw} N.~Seiberg and E.~Witten, \ct{String theory and
noncommutative geometry} \jhep{09}{1999}{032}; \phepth{9908142}.

\bibitem{dnrev} M.~R.~Douglas and N.~A.~Nekrasov, \ct{Noncommutative
Field Theory}
to appear in {\ifuseprd\else\it\fi Rev.\ Mod.\ Phys.},
\hepth{0106048}.

\bibitem{liu} H.~Liu, \ct{$\ast$-Trek II: $\ast_n$ Operations, Open Wilson
Lines and the Seiberg-Witten Map}
\hepth{0011125}.

\bibitem{lm3} H.~Liu and J.~Michelson, 
\ct{Supergravity Couplings of Noncommutative D-branes},
RUNHETC-2001-01,
\hepth{0101016}.

\bibitem{lm4} H.~Liu and J.~Michelson, 
\ct{Ramond-Ramond Couplings of Noncommutative D-branes},
RUNHETC-01-12, NSF-ITP-01-29,
\hepth{0104139}.

\bibitem{dastrivedi} S.~R.~Das and S.~P.~Trivedi, \ct{Supergravity
Couplings to Noncommutative Branes, Open Wilson Lines and Generalised
Star Products}, \jhep{02}{2001}{046}; \phepth{0011131}; \\
S.~R.~Das, \ct{Bulk couplings to noncommutative branes} \hepth{0105166}.


\bibitem{ooguri} Y.~Okawa and H.~Ooguri, \ct{How Noncommutative Gauge
Theories Couple to Gravity} \npb{599}{2001}{55--82}; \phepth{0012218}; \\
Y.~Okawa and H.~Ooguri, \ct{Energy-momentum tensors in Matrix theory and 
in noncommutative gauge  theories} \hepth{0103124}.


\bibitem{oo} Y.~Okawa and H.~Ooguri, \ct{An Exact Solution to
Seiberg-Witten Equation of Noncommutative Gauge Theory}
CALT-68-2325, CITUSC/01-010, NSF-ITP-01-25, \hepth{0104036}.

\bibitem{mukhi} S. Mukhi and N.~V.~Suryanarayana, \ct{Chern-Simons
Terms on Noncommutative Branes} \jhep{11}{2000}{006},
\phepth{0009101};\\
S.~Mukhi and N.~V.~Suryanarayana, \ct{Ramond-Ramond couplings of 
noncommutative branes} 
TIFR/TH/01-24, \hepth{0107087}.

\bibitem{mukhi2} S.~Mukhi and N.~V.~Suryanarayana, \ct{Gauge-Invariant
Couplings of Noncommutative Branes to Ramond-Ramond Backgrounds} 
\jhep{01}{2001}{023};
\phepth{0104045}.

\bibitem{garousi1} M.~R.~Garousi, \ct{Dirac-Born-Infeld action, 
Seiberg-Witten map, and Wilson lines}
IPM/P-2001/009, \hepth{0105139}; \\
M.~R.~Garousi, \ct{Transformation of the 
Dirac-Born-Infeld action under the 
Seiberg-Witten map} \npb{602}{2001}{527}; \phepth{0011147}.

\bibitem{dasmu} S.~R.~Das, S.~Mukhi and N.~V.~Suryanarayana,
\ct{Derivative corrections from noncommutativity} \hepth{0106024}.

\bibitem{dhar} A.~Dhar and Y.~Kitazawa,
\ct{Non-commutative gauge theory, open Wilson lines and closed strings}
\hepth{0106217}.

\bibitem{iikk} N.~Ishibashi, S.~Iso, H.~Kawai, Y.~Kitazawa,
\ct{Wilson Loops in Noncommutative Yang Mills} 
\npb{573}{2000}{573--593}; \phepth{9910004}.
 
\bibitem{ambjorn} J.~Ambjorn, Y.~M.~Makeenko, J.~Nishimura and 
R.J.~Szabo,
\ct{Finite N Matrix Models of Noncommutative Gauge Theory} 
\jhep{11}{1999}{029}; \phepth{9911041}; \\
J.~Ambjorn, Y.~M.~Makeenko, J.~Nishimura and
R.J.~Szabo,
\ct{Lattice Gauge Fields and Discrete Noncommutative Yang-Mills
Theory}
\jhep{05}{2000}{023}; \phepth{0004147}.
  
\bibitem{dasrey} S.~R.~Das and S.-J.~Rey, \ct{Open Wilson Lines in
Noncommutative Gauge Theory and Tomography of Holographic Dual
Supergravity} 
\npb{590}{2000}{453--470}; \phepth{0008042}; \\
S.~Rey and R.~von Unge,
\ct{S-duality, noncritical open string and noncommutative gauge theory}
\plb{499}{2001}{215}; \phepth{0007089}.

\bibitem{ghi} D.~J.~Gross, A.~Hashimoto and N.~Itzhaki,
\ct{Observables of Non-Commutative Gauge Theories} NSF-ITP-00-94,
\hepth{0008075}.

\bibitem{bl} D.~Berenstein and R.~G.~Leigh, \ct{Observations on
non-commutative field theories in coordinate space}
ILL-(TH)-00-11; \hepth{0102158}.

\bibitem{kiemrey} Y.~Kiem, S.~Rey, H.~Sato and J.~Yee,
\ct{Open Wilson lines and generalized star product in nocommutative 
scalar  field theories} \hepth{0106121}; \\
Y.~Kiem, S.~Rey, H.~Sato and J.~Yee, \ct{Anatomy of One-Loop Effective 
Action in Noncommutative Scalar Field Theories} \hepth{0107106}.


\bibitem{garousi} M.~R.~Garousi, \ct{Non-commutative world-volume
interactions on D-brane and Dirac-Born-Infeld action}
\npb{579}{2000}{209--228}; \phepth{9909214}.


\bibitem{lm2} H.~Liu and J.~Michelson, \ct{$\ast$-TREK: The 
One-Loop N=4 Noncommutative SYM Action} \hepth{0008205}.

\bibitem{mehen} T.~Mehen and M.~B.~Wise, \ct{Generalized $\star$-Products, 
Wilson Lines and the Solution of the Seiberg-Witten Equations} 
\jhep{12}{2000}{008};
\phepth{0010204}.

\bibitem{bs} D.~Bigatti and L.~Susskind, \ct{Magnetic Fields, Branes 
and Noncommutative Geometry} 
\citeprd{62}{2000}{066004}; \phepth{9908056}.
 
\bibitem{sj} M.~M.~Sheikh-Jabbari, \ct{Open Strings in a B-field
Background as Electric Dipoles} \plb{455}{1999}{129--134}; 
\phepth{9901080}.

\bibitem{yin} Z.~Yin, \ct{A Note on Space Noncommutativity}
\plb{466}{1999}{234--238}; \phepth{9908152}.

\bibitem{elliott} G. A. Elliott, \ct{On the K-theory of the 
$C^{\ast}$-algebra generated by a projective representation of a 
torsion-free discrete abelian group} in {\em Operator Algebras 
and Group Representations}, Vol 1. 1984  Pitman, London.

\bibitem{rieffel} M. Rieffel, \ct{Projective modules over higher-dimensional
noncommutative tori} \cjm{40}{1988}{257--338}.

\bibitem{schwarz} A. Schwarz, \ct{Morita equivalence and duality} 
\npb{534}{1998}{720-738};\phepth{9805034}.

\bibitem{schwarzre} A. Konechny and  A. Schwarz, \ct{Introduction to 
M(atrix) theory and noncommutative geometry} \hepth{0012145}.

\bibitem{hi} A.~Hashimoto and N.~Itzhaki, \ct{Non-Commutative 
Yang-Mills and the AdS/CFT Correspondence} \plb{465}{1999}{142--147};
\phepth{9907166}.
 
\bibitem{mr} J.~M.~Maldacena and J.~G.~Russo, \ct{Large N Limit of 
Non-Commutative Gauge Theories} \jhep{09}{1999}{025}; 
\phepth{9908134}.

\bibitem{divech} M.~Bill\'{o}, P.~Di Vecchia, M.~Frau, A.~Lerda,
I.~Pesando, R.~Russo and  S.~Sciuto,
\ct{Microscopic string analysis of the D0-D8 brane system and dual R-R
states}
\npb{526}{1998}{199--228}; \phepth{9802088}.

\bibitem{jp} J.~Polchinski, \bt{String Theory, Vol. I and II} (Cambridge
University Press,
Cambridge, England, 1998).

\bibitem{callan} C.~G.~Callan, C.~Lovelace, C.~R.~Nappi and
S.~A.~Yost, 
\ct{Loop Corrections to Superstring Equations of Motion}
\npb{308}{1988}{221--284}.

\bibitem{berg} E.~Bergshoeff and P.~K.~Townsend, \ct{Super D-branes} 
\npb{490}{1997}{145}; \phepth{9611173}.

\bibitem{gr} I.~S.~Gradshteyn and I.~M.~Rhyzhk, \bt{Table of
Integrals, Series, and Products, Fifth Edition} Academic Press,
Inc., San Diego, CA, 1994.


\bibitem{bmz} D. Brace, B. Morariu and B. Zumino, \ct{T-Duality and 
Ramond-Ramond Backgrounds in the Matrix Model} \npb{549}{1999}{181-193};
\phepth{9811213}.

\bibitem{fot} M.~Fukuma, T.~Oota and H.~Tanaka,
\ct{Comments on T-dualities of Ramond-Ramond Potentials}
\ptp{103}{2000}{425--446}; \phepth{9907132}.

\bibitem{seiberg} N.~Seiberg, \ct{A Note on Background Independence in
Noncommutative Gauge Theories, Matrix Model and Tachyon Condensation}
\jhep{09}{2000}{003}; \phepth{0008013}.

\bibitem{wadia} A.~Dhar, and S.~R.~Wadia, \ct{A Note on Gauge Invariant 
Operators in Noncommutative Gauge Theories and the Matrix Model} 
\plb{495}{2000}{413--417},
\phepth{0008144}.

\bibitem{gm2} M.~R.~Garousi and R.~C.~Myers,
\ct{World-Volume Interactions on D-branes} \npb{542}{1999}{73--88},
\phepth{9809100}; \\
M.~R.~Garousi and R.~C.~Myers,
\ct{World-Volume Potentials on Branes} \jhep{11}{2000}{032}, \phepth{0010122}.

\bibitem{oow} H.~Ooguri, private communication.

\bibitem{malm} S.~S.~Gubser, A.~Hashimoto, I.~R.~Klebanov and J.~M.~Maldacena,
\ct{Gravitational lensing by $p$-branes}
\npb{472}{1996}{231--248}; \phepth{9601057}.

\bibitem{myersm} M.~R.~Garousi and R.~C.~Myers, \ct{Superstring
Scattering from D-Branes} \npb{475}{1996}{193--224}; \phepth{9603194}.

\bibitem{witd0} E.~Witten, \ct{BPS bound states of D0-D6 and D0-D8 
systems in a B-field} \hepth{0012054}.

\end{thebibliography}
\end{document}